\documentclass[10pt]{article}
\usepackage[OE]{express}
\usepackage{color}
\usepackage{amsmath,subfigure}

\begin{document}

\title{Dynamics of a broad-band Quantum Cascade Laser. From chaos to coherent dynamics and mode-locking}

\author{L. L. Columbo$^{1,2*}$, S. Barbieri $^{3,4}$, C. Sirtori $^{3}$, and M. Brambilla $^{2,5}$}

\address{
$^1$ Dipartimento di Elettronica e Telecomunicazioni, Politecnico di Torino, Corso Duca degli Abruzzi 24, 10129 Torino, Italy\\
$^2$ Consiglio Nazionale delle Ricerche, CNR-IFN, via Amendola 173, 70126 Bari, Italy\\
$^3$ Laboratoire Mat\'eriaux et Ph\'enom$\grave{e}$nes Quantiques, Universit\'e Paris Diderot and CNRS, UMR 7162,
10 rue A. Domont et L. Duquet, F-75205 Paris, France\\
$^4$ Institut d'Electronique, de Micro\'electronique et de Nanotechnologie, UMR CNRS 8520, Avenue Poincar\'e, 59652 Villeneuve d?Ascq, France
$^5$ Dipartimento Interateneo di Fisica, Universit$\grave{a}$ degli Studi e Politecnico di Bari, via Amendola 173, 70126 Bari, Italy}

\email{*lorenzo.columbo@polito.it} 


\begin{abstract}
The dynamics of a multimode Quantum Cascade Laser, is studied in a model based on effective semiconductor Maxwell-Bloch equations, encompassing key features for the radiation-medium interaction such as an asymmetric, frequency dependent, gain and refractive index as well as the phase-amplitude coupling provided by the Henry factor.
By considering the role of the free spectral range and Henry factor, we develop criteria suitable to identify the conditions which allow to destabilize, close to threshold, the traveling wave emitted by the laser and lead to chaotic or regular multimode dynamics. In the latter case our simulations show that the field oscillations are associated to self-confined structures which travel along the laser cavity, bridging mode-locking and solitary wave propagation. In addition, we show how a RF modulation of the bias current leads to \textit{active} mode-locking yielding high-contrast, picosecond pulses. Our results compare well with recent experiments on broad-band THz-QCLs and may help understanding the conditions for the generation of ultrashort pulses and comb operation in Mid-IR and THz spectral regions.
\end{abstract}

\ocis{(000.0000) General.} 


\section{Introduction}

The multimode dynamics of Mid-IR and THz Quantum Cascade Lasers (QCLs) became a focus of interest for the realization of pulsed regimes and frequency combs for a number of applications including
time-resolved measurements, frequency mixing, high-precision spectroscopy etc. \cite{faist}-\cite{Barbieri2017}. Such phenomena can be retraced to the capability of realizing coherent locking of modes in a multimode emission regime.\\
In literature, different models have been proposed to theoretically interpret the experimental evidences \cite{Gordon}-\cite{Tzenov}. A first class of models are essentially based on Maxwell-Bloch equations (MBEs) ($in 2$ or $3$-level medium approximation) \cite{Gordon, capasso0, capasso1, Kurgin, villares, Boiko, Tzenov}, where in some cases additional effects (e.g. cubic nonlinearity, saturable absorption, cavity dispersion) have been phenomenologically introduced to provide a proper interpretation of the experimental findings. A common understanding emerging thereof was the role of Spatial Hole Burning (SHB) (introduced to describe the effect of counter--propagating fields in a Fabry--Perot (FP) scheme of the laser cavity) in reducing the instability threshold leading to the mode-locking (see e.g. \cite{Boiko}).
In other approaches adopting a closer semiconductor (s.c.) optical response \cite{erneux, elsasser}, a standard adiabatic elimination of the macroscopic semiconductor polarization has been introduced, leading to the commonly termed \textit{rate equation} model. While this approximation proved successful in describing the dynamics of (especially single mode) optical nonlinear systems, it is known that it entails unphysical dynamical effects in multimode lasers because it corresponds to an infinitely broad gain/dispersion \cite{Oppo}.
In order to circumvent this limitation, spectral filtering terms were added in the description of broad-band QCLs \cite{elsasser}.
On a more fundamental basis, other works report on nonstandard techniques for the adiabatic elimination of the polarization which formally amount to introducing diffusive terms in the equations for the multimode laser field \cite{moloney}.
In order to provide a fundamentally self--consistent modelization, it is desirable to take into account a frequency dependent, asymmetric dispersion/gain line and the Linewidth Enhancement Factor (LEF) that are known to play a critical role in coherent multimode regimes of s.c. lasers \cite{Prati,Prati1}.\\
Thus, in this work we adapt to a QCL, a s.c. laser model whose basics have already been proved quite effective in the description of multimode instabilities in bipolar lasers \cite{Prati, Prati1, Gustave}. Also, in order to reduce the complexity of the model, we retain the transition dynamics in the 2-level framework \cite{faist} \footnote{To this end we note that a proof of the reduction of three level dynamics, probably more suitable for a QCL, to an effective-two-level scheme was provided in \cite{pessina} in the case of fiber-lasers.}. Specifically, we introduce an asymmetric gain/dispersion line in the linear-gain approximation fitting the gain peak position and width to the experimental data and account for the LEF without resorting to first principle derivation of the full polarization evolution.\\
In the model, we consider a unidirectional ring cavity configuration;
this hypothesis allows us to show that phase-amplitude coupling (associated to the LEF) and mode competition can provide, even in absence of SHB typical of Fabry-Perot (FP) configurations considered e.g. in \cite{Gordon, Boiko}, a radically lower threshold for coherent multimode dynamics than what predicted by the Risken-Nummedal-Graham-Haken (RNGH) instability \cite{RN, GH, NOS}, which is typical of two-levels laser.
In this regard, we observe that unidirectional lasing in ring cavities in QCLs has been demonstrated in both a monolithically integrated cavity and an external one \cite{Sorel, Malara}. Recently, in the latter configuration, working in both unidirectional and bi-directional emission regimes, the first experimental evidence of active mode-locking in QCLs has been reported \cite{Capasso2016}. On a farther perspective we might envisage that compact QCL ring resonators would share a similar future with the Near-IR ring resonators that nowadays represent key elements in integrated photonics circuits \cite{Liu}. \\
The extension of the present model to a FP resonator, where other physical mechanisms such as SHB may play a role in the system dynamics and possibly hinder the formation of stable pulses via mode-locking \cite{Malara,capasso1} is left to a future work. \\

In Section \ref{sec2} we (i) introduce the s.c. laser model, (ii) derive an analytical expression for the singlemode laser solutions in the free-running regime in the form of traveling waves (TWs) and (iii) study their respective thresholds when the LEF and free spectral range (FSR) are varied. As expecEFd,
the TW closest to the gain peak has the lowest threshold, but at higher pump currents, we predict competition among several laser modes. In this case a "rule of thumb" can be obtained to gauge the laser operating regimes towards irregular (spatio-temporal chaos) or coherently locked emission, the latter leading to regular oscillations of the output field. \\
In Section \ref{sec3} our simulations show that increasing the bias current leads to the destabilization of the lowest threshold TW and drives the laser through a sequence of alternating irregular regimes and regular, mode-locked, regimes in agreement with recent experiments results on the multimode dynamics of an ultra-broad-gain THz-QCL \cite{Scalari1, barbieri}. Also, we find that a larger LEF (more typical of a Mid-IR QCL \cite{faist}) favors the TW destabilization towards multimode regimes. \\
In the regular, mode-locked regimes we interpret the pulsed dynamics with the formation inside the cavity of self--confined structures, with a variable number of intensity peaks, which can also explain the experimentally observed disappearance of the fundamental beat note (BN) in the spectrum.
%
%
Such structures have a clear solitonic character, not unlike those recently observed in long cavity (i.e. "quasi class-A") s.c.lasers \cite{Gustave}. They appear as low-contrast pulsed emission on a CW background and therefore represent an interesting example of a \textit{spontaneous mode-locking} in a QCL laser.\\
While such a small peak contrast is a limitation for applications, in Section \ref{sec4} we show the generation of picosecond pulses, introducing a RF modulation in a portion of the cavity which might have a relevant impact on applications such as time resolved measurements, high-precision spectroscopy, multispectral sensing and imaging. \\
In Section \ref{concl} we summarize our main results and trace the pathways to future investigations.

\section{Effective Semiconductor Bloch Equations}\label{sec2}

\begin{figure}[htb]
\begin{center}
\subfigure[]{\includegraphics[width=0.5\textwidth]{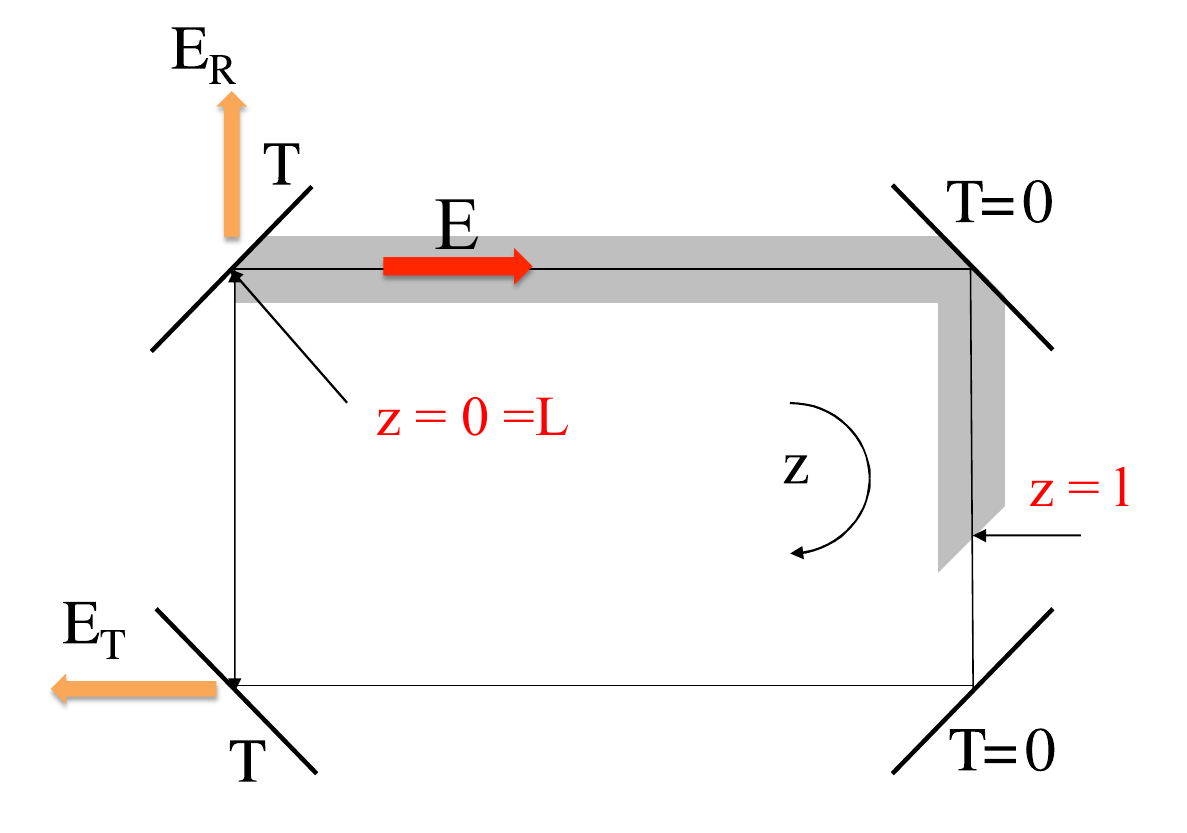}}
\subfigure[]{\includegraphics[width=0.4\textwidth]{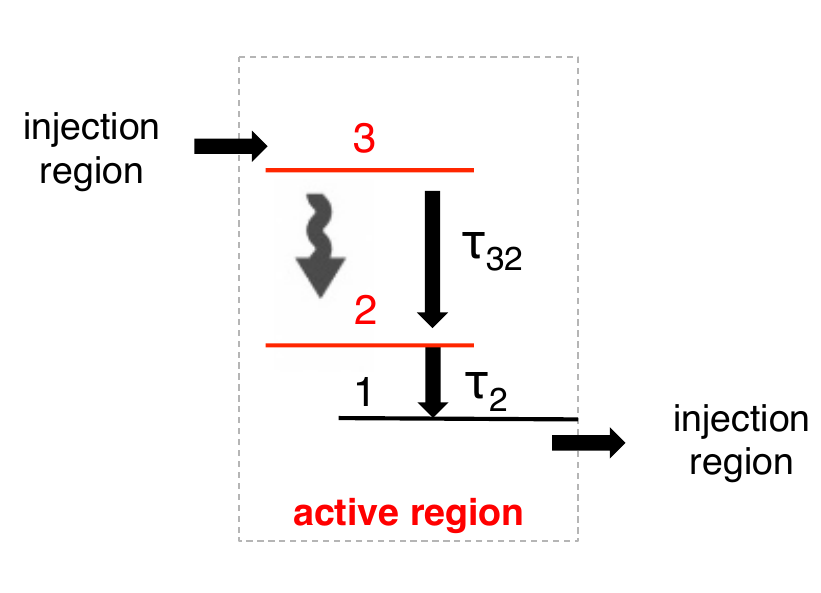}}
\end{center}
\caption{(a) Sketch of a semiconductor laser in a unidirectional ring cavity. (b) Schematic conduction band diagram of a QCL. Each stage of the structure consists in an active region and a injection region. Lasing action occurs between level $3$ and level $2$. Thus, the lifetime of the nonradiative transition $3$ $\longrightarrow$ $2$ denoted as $\tau_{32}$ has to be longer than the lifetime $\tau_{2}$ of the level $2$. The lifetime of the level $1$ is supposed to be $0$.}
\label{fig4}
\end{figure}

We consider the ring resonator sketched in Fig. \ref{fig4}.a where a semiconductor active medium of length $l \sim 1 $ $mm$ is placed in a unidirectional ring cavity of total length $L\ge l$. We suppose that two mirrors have transmissivity $T \ne 0$, while the others are perfect reflectors ($T=0$). This configuration choice excludes the onset of the spatial hole burning multimode instability \cite{capasso1}.\\
In the hypothesis of linearly polarized field, we write
\begin{equation}
\tilde{E}(z,t)=\frac{E(z,t)}{2}\exp{\left[i(k_{0}z-\omega_{0}t)\right]}+c.c.
\end{equation}
where $k_{0}=\omega_{0}/v=\omega_{0}n/c$, $n=\sqrt{\epsilon_{b}}$ is the background refractive index and $\omega_{0}$ is the angular frequency of an empty cavity mode and it will be taken in the following as the reference frequency. \\
Analogously, the medium macroscopic polarization can be written as:
 \begin{equation}
\tilde{P}(z,t)=\frac{{P(z,t)}}{2}\exp{\left[i(k_{0}z-\omega_{0}t)\right]}+c.c.
\end{equation}
In the slowly varying envelope and rotating wave approximation, the radiation-matter interaction is then described by the following nonlinear partial differential equations:
\begin{eqnarray}
\frac{\partial E}{\partial z}+\frac{1}{v}\frac{\partial E}{\partial t} &=& g\, P \label{E}\\
\frac{\partial N}{\partial t}&=& \frac{I}{eV}-\frac{N}{\tau_{e}}-\frac{i}{4 \hbar}(E^{*}P-EP^{*})\label{P}
\end{eqnarray}
where $N$ is the carrier density in the upper laser level (in the hypothesis of instantaneous relaxation of the lower state ($\tau_{2}=0$) \cite{faist,capasso0, capasso1}), $\tau_{e}=\tau_{32}$ is the carrier density nonradiative decay time, $I$ and $V$ are the pump current and the sample volume respectively, $g=\frac{i N_{p}\omega_{0}\Gamma_{c}}{2 \epsilon_{0}n c}$, $\Gamma_{c}$ is the overlap factor between the optical mode and the active region and $N_{p}$ is the number of cascading gain stages.
The Fourier components of $P$ are proportional to the intracavity field via a complex susceptibility:

$$\hat{P}(\omega)= \epsilon_{0}\epsilon_{b}\chi(\omega)\hat{E}(\omega)$$

which can be referred to the resonant medium in a general way by assuming;
\begin{equation}
\chi(\omega,N)=\frac{A(N)}{B(N)-i\omega}  \label{chi1}
\end{equation}
so that, anti-transforming, we get the following dynamical equation for $P$:
\begin{equation}
\frac{\partial P}{\partial t}= \epsilon_{0}\epsilon_{b}A(N)\,E-B(N)\, P
\end{equation}
By considering complex quantities $A=Re(A)+iIm(A)$ and $B=Re(B)+iIm(B)$, the susceptibility takes the form:
\begin{figure}
\begin{center}
\subfigure[]{\includegraphics[width=0.4\textwidth]{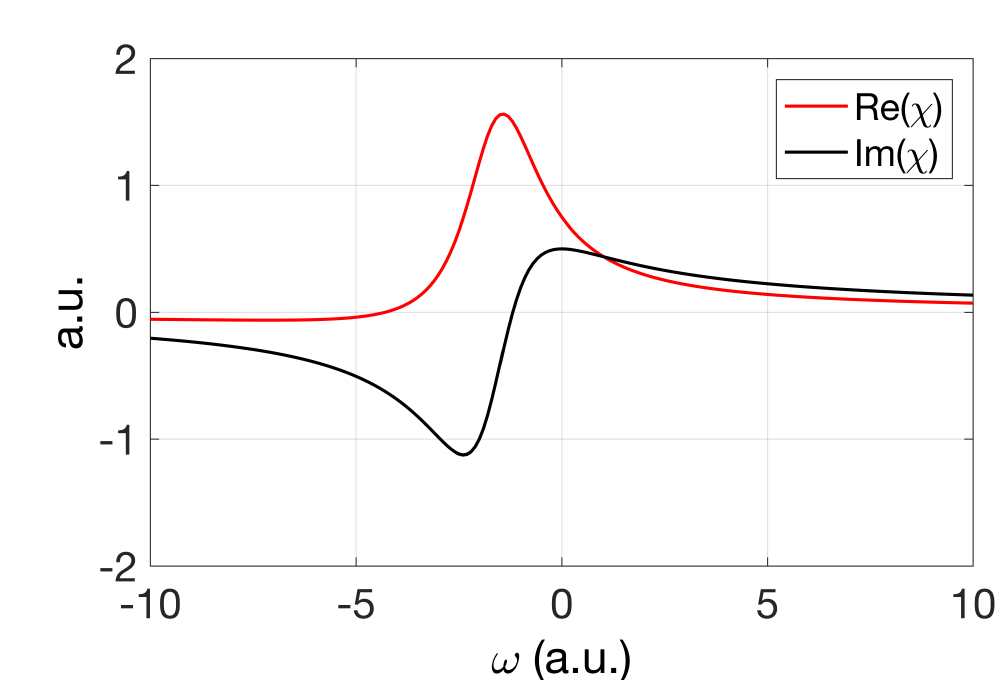}}
\subfigure[]{\includegraphics[width=0.4\textwidth]{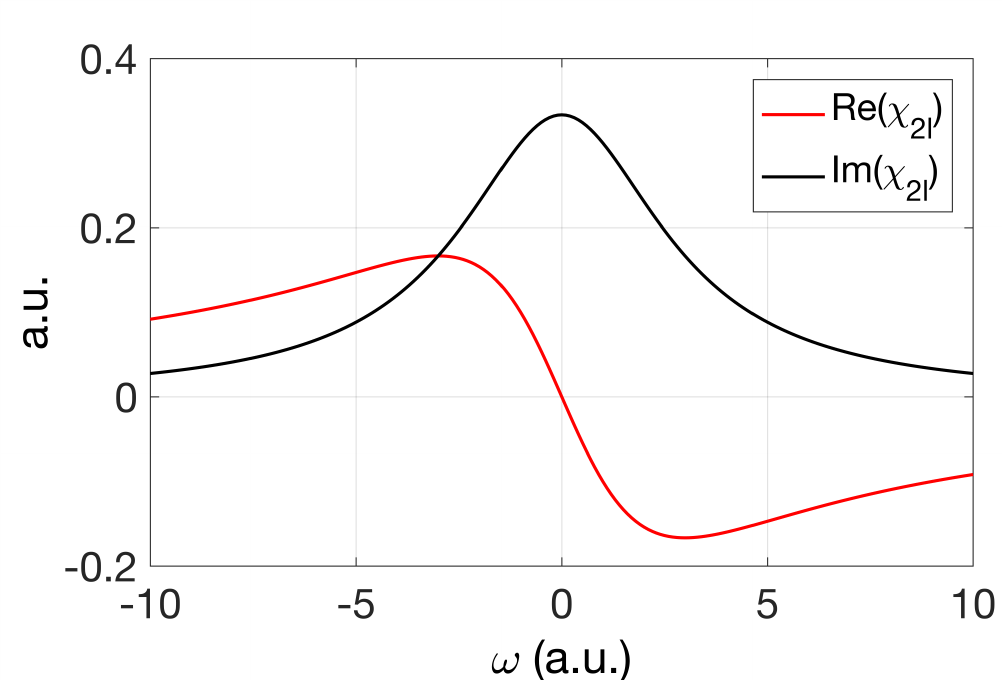}} 
\end{center}
\caption{(a) Phenomenological s.c. susceptibility. (b) Two-level system susceptibility.} \label{susc}
\end{figure}

\begin{equation}
\chi=\frac{Re(A)Re(B)+Im(A)(Im(B)-\omega)}{Re(B)^{2}+(Im(B)-\omega)^{2}}+i \frac{Im(A)Re(B)-Re(A)(Im(B)-\omega)}{Re(B)^{2}+(Im(B)-\omega)^{2}} \label{chi2}
\end{equation}
Note that it must be $Re(B)>0$ in order for $\chi$ to represent an analytical function of $\omega$ in the upper half plane, and in particular we will take
\begin{equation}
B=\left(\Gamma+i\delta \right)/ \tau_{d}, \, (\Gamma\in \mathbf{R^+} \, , \, \delta \in \mathbf{R})    \label{B}
\end{equation}
where $\tau_{d}$ is the dipole dephasing time and $\Gamma$ and $\delta$ are reasonably assumed independent from $N$ \cite{Prati}.\\
As illustrated by Fig.\ref{susc}.a, the (asymmetric) gain and the refractive index curves associated to $Re(\chi)$ and $Im(\chi)$ now lack even or odd parity and moreover, the zero of the latter does not coincide with the maximum of the former. The the canonically symmetrical dispersion curve and Lorentzian gain of the two--level model can be readily recovered by assuming $Re(A)=0$. i.e.:
\begin{equation}\chi_{1,2lev}=\frac{Im(A)(Im(B)-\omega)}{Re(B)^{2}+(Im(B)-\omega)^{2}}+i \frac{Im(A)Re(B)}{Re(B)^{2}+(Im(B)-\omega)^{2}} \label{chi2lev} \end{equation}
Figure.\ref{susc}.b evidences this difference. \\
From Eq.(\ref{chi2}) it follows that the maximum of $Im(\chi)$, and hence of the gain curve is given by:
\begin{equation}
\omega_{M}=Im(B)-Re(B)\left( \frac{Im(A)}{Re(A)}-\sqrt{\left(\frac{Im(A)}{Re(A)}\right)^{2}+1} \right) \label{maxeq}
\end{equation}
Therefore, in the limit $Im(A)/Re(A)>>1$, (or in the two-level case, $Re(A)=0$), one has $\omega_{M} \simeq \delta/\tau_{d}$ and the gain linewidth (FWHM) is approximately $\Gamma/\tau_{d}$, so that we recover the intuitive picture for the susceptibility behavior (Fig.\ref{susc}.b). \\
A further simplification can be introduced for moderate carrier densities (as it will be valid in the following sections), by modeling a linear dependence of the s.c. susceptibiity on $N$ at the reference frequency (see e.g. Eq. (2.115) of \cite{chow} as follows:
\begin{equation}
\frac{A(N)}{B}=-f_{0}(\alpha+i)N, (f_{0} \in \mathbf{R}) \label{A}
\end{equation}
where $\alpha$ is the LEF. Hence using Eq.(\ref{A}) we get:
\begin{equation}
A(N)=-f_{0}N  \left( \frac{\alpha \Gamma -\delta}{\tau_d}  \right) -i\, f_{0}N  \left( \frac{\Gamma +\alpha\delta}{\tau_d}  \right) \label{Aparam}
\end{equation}
Setting $\delta=-\alpha \Gamma$ (so that $Re(A)$ $=$ $-2f_{0}N \alpha \Gamma/\tau_d$) implies the reasonable assumption that the gain maximum coincides with the reference cavity mode (from Eq. (\ref{maxeq}): $\omega_{M}=0$ ). In this way we recover the two-level description by setting $\alpha=0$.
%
The real constants $f_{0}$, $\alpha$, $\Gamma$ can be obtained by fitting the experimentally measured gain spectra reported for example in \cite{Scalari1,Scalari2} around their maxima \cite{Prati}.\\
By using relations (\ref{B}) and (\ref{A}) in Eqs. (\ref{E})-(\ref{P}) we get the Effective Semiconductor Maxwell-Bloch Equations (ESMBEs) \cite{Ning} which correctly incorporate the peculiarities of the s.c. laser nonlinearities:
 \begin{eqnarray}
\frac{\partial E}{\partial z}+\frac{1}{v}\frac{\partial E}{\partial t} &=& g\, P \label{E1}\\
\frac{\partial P}{\partial t}&=& \frac{1}{\tau_{d}}\Gamma \left(1-i \alpha \right)\left[-P-if_{0}\epsilon_{0}\epsilon_{b}(1-i\alpha) E\, N   \right]   \\
\frac{\partial N}{\partial t}&=&\frac{1}{\tau_{e}}\left[ \frac{I\tau_{e}}{eV}-N-\frac{i \tau_{e}}{4 \hbar}(E^{*}P-EP^{*})\right] \label{P1}
\end{eqnarray}

We also remark that (1)
the adiabatic elimination of the macroscopic polarization $P$ obtained by setting $\frac{\partial P}{\partial t}=0$ in Eqs. (\ref{E1})-(\ref{P1}) corresponds to assuming a flat gain curve and it leads to the rate equation model analogous to that adopted in \cite{erneux} (in the three--level case) and (2)
the adiabatic elimination of the macroscopic polarization $P$ supplemented by the inclusion of a spectral filtering term in Eq.(\ref{E1}) to model the finite gain linewidth, leads to a modified rate equation model as adopted in \cite{elsasser} (in the three--level case).\\

A more compact form can be achieved by introducing the following variable changes and scaled parameters:
$$\eta_{1}E \longrightarrow E, \quad i\eta_{2}P\longrightarrow P,\quad \eta_{3}N\longrightarrow D$$
$$\eta_{1}^{2}=\frac{\eta_{3}\tau_{e}}{2 \hbar}, \quad \eta_{2}=\eta_{1}, \quad \eta_{3}=\epsilon_{0}f_{0}\epsilon_{b}, \quad \tilde{g}=-ig \, \in \mathbf{R}, \quad \mu=\frac{I\eta_{3}\tau_{e}}{eV}$$
so that Eqs.(\ref{E1})-(\ref{P1}) become:
 \begin{eqnarray}
\frac{\partial E}{\partial z}+\frac{1}{v}\frac{\partial E}{\partial t} &=& \tilde{g} P \label{E21}\\
\frac{\partial P}{\partial t}&=& \frac{1}{\tau_{d}} \Gamma \left(1-i \alpha \right)\left[(1-i\alpha)E\, D-P \right]   \\
\frac{\partial D}{\partial t}&=&\frac{1}{\tau_{e}}\left[ \mu-D-\frac{1}{2}(E^{*}P+E P^{*})\right] \label{P21}
\end{eqnarray}
%
%
%
%
The final formal steps are (i) the introduction of the boundary conditions modelling the ring cavity, which allows us to isolate effects of s.c. medium and mode spectra from SHB due to standing wave gratings,
(ii) the application of the low transmission limit, which makes the concept of \textit{cavity mode} well defined, and (iii) the assumption that the medium completely fills the cavity ($L=l$ in Fig.1a) which is coherent with the focus on millimetric monolithic devices, although our model is apt to describe extended cavity configurations as those in \cite{capasso1, Capasso2016}. \\
Since the procedure is rather standard \cite{NOS}, we refer the reader to Appendix A where Eqs. (\ref{E21})-(\ref{P21}) are worked in the form:\\
\begin{eqnarray}
\frac{c \tau_{d}}{ln}\frac{\partial \tilde{E}}{\partial z}+\frac{\partial \tilde{E}}{\partial t} &=&\sigma \left(-\tilde{E}+\tilde{P}\right) \label{E4b}\\
\frac{\partial \tilde{P}}{\partial t}&=& \Gamma \left(1-i \alpha \right)\left[(1-i\alpha)\tilde{E} \tilde{D}-\tilde{P} \right]   \label{P4b}\\
\frac{\partial \tilde{D}}{\partial t}&=&b\left [ \tilde{\mu}-\tilde{D}-\frac{1}{2}\left(\tilde{E}^{*}\tilde{P}+ 	\tilde{E} \tilde{P}^{*} \right)\right]\label{P4b1}
\end{eqnarray}
where $\tilde{E}=E$, $\tilde{P}=\mathcal{A} P$, $\tilde{D}=\mathcal{A} D$ and $\tilde{\mu}=\mathcal{A}\mu$ with $\mathcal{A}=\tilde{g}l/T$. Time is normalized to the fastest timescale $\tau_{d}$ so that $\sigma=\tau_{d}/\tau_{p}$, $b=\tau_{d}/\tau_{e}$ with $\tau_{p}=nl/cT$, and the longitudinal coordinate to the cavity length $l$. \\
Although our results might also apply to Mid-IR-QCLs, we adopt in the following typical values for a THz-QCL by setting $\tau_{d}=0.1ps$, $\tau_{p}=10ps$, $\tau_{e}=1ps$ (that give $\sigma=0.01$ and $b=0.1$). The considered value of $\tau{p}$ is corrected to account for the total losses inside the laser cavity \cite{villares}.
%
%
\subsection{Free running traveling wave solutions}
Due to the peculiarities of the s.c. laser we considered, the stationary solutions differ in intensity and frequency from the known ones pertaining to 2--level lasers. The solution is sought among TWs of longitudinal wavevector $k$ and frequency $\Omega$ in the form:
\begin{equation}
\tilde{E}=E_{s}e^{i(kz-\Omega t)} \quad \tilde{P}=P_{s}e^{i(kz-\Omega t)} \quad \tilde{D}=D_{s} \label{CW}
\end{equation}
Setting to zero the RHS of Eqs. (\ref{E4b})-(\ref{P4b1}), using expressions (\ref{CW}) and dropping the tilde we have:
\begin{eqnarray}
\frac{c \tau_{d}}{ln}ikE_{s}-i \Omega E_{s} &=&\sigma \left(-E_{s}+P_{s}\right) \label{E4c}\\
-i \Omega P_{s}&=& \Gamma\left(1-i \alpha \right)\left[(1-i\alpha)E_{s} D_{s}-P_{s} \right]   \label{P4c}\\
0&=&\mu-D_{s}-\frac{1}{2}\left(E_{s}^{*}P_{s}+ E_{s} P_{s}^{*} \right) \label{Dc}
\end{eqnarray}
and hence the stationary solutions for $P_{s}$ and $D_{s}$ are:
\begin{eqnarray}
P_{s}&=&E_{s}\frac{\mu}{ (1+|E_{s}|^{2}G_{1})}(G_{1}+iG_{2})  \\
D_{s}&=&  \frac{\mu}{ (1+|E_{s}|^{2}G_{1}) }
\end{eqnarray}
where
$$G_{1}=\left(\Gamma^{2}-\alpha^{2} \Gamma^{2}\right)/\left[\Gamma^{2}+(\alpha \Gamma+\Omega)^{2})\right
],$$
$$G_{2}=\left[-2\alpha \Gamma^{2} +\Gamma(\alpha \Gamma+\Omega)-\alpha^{2} \Gamma (\alpha \Gamma+\Omega)\right]/\left[\Gamma^{2}+(\alpha \Gamma+\Omega)^{2})\right]$$
The stationary field solution from Eq. (\ref{E4c}) is then:
\begin{equation}
E_{s}\left(\frac{c \tau_{d}}{ln}ik-i\Omega \right)=\sigma E_{s} \left[-1+ \frac{\mu(G_{1}+iG_{2})}{ (1+|E_{s}|^{2}G_{1}) }\right] \\
\end{equation}
Two real equations follow, which constitute the generalization of the stationary lasing intensity and the (implicit) dispersion relation $\Omega(k)$ of the TW angular frequency for the considered QCL:
\begin{eqnarray}
|E_{s}|^{2}&=&\mu -\frac{1}{G_{1}} \label{field}\\
\Omega&=&\frac{c \tau_{d}}{ln}k - \sigma \frac{G_{2}}{G_{1}} \label{freq}
\end{eqnarray}

\subsection{Traveling wave selection at threshold }

A general stability analysis for the TW solutions derived above is left to a future work, while in this instance we will focus on the destabilization of the TW emitted at threshold and to the multimode regimes appearing beyond it. In particular we are interested in the role played by the mode separation and the LEF. The instability of the TW solutions in QCL models based on MBEs for 2--level systems in case of FP and ring configurations has been linked to the RNGH instability, associated with the parametric gain of the cavity modes in resonance with the Rabi frequency \cite{Gordon,Boiko}. In our case, though, the phase--amplitude coupling points towards a different character, namely the Benjamin--Feir instability \cite{BFI}, that appears to be peculiar of the multimode s.c. laser dynamics as assessed in \cite{Prati, gil} and is triggered by the growth of modes having wavevectors larger than a critical value that depends on boundary conditions. In particular in \cite{gil} the authors, using to a codimension $2$ bifurcation analysis and the hypothesis of instantaneous medium response ("class A" laser), show how a set of effective semiconductor Bloch equations analogous to those used here can be reduced close to threshold to the prototypical cubic Complex Ginzburg Landau Equation (CGLE) that describes the behavior of a large class of spatially extended nonlinear systems \cite{Aranson}. The "phase instability" (or Benjamin--Feir instability) of the TW solutions of the CGLE leads to a rich variety of spatio-temporal complexity encompassing phase-- and defect--mediated turbulence (where respectively small or large amplitude modulations of the unstable TW occur) and coherent phenomena as the formation of localized structures. Because of the small carriers relaxation time that makes QCLs "quasi class A" lasers, we may thus expect to observe a similar dynamical scenario. We note that with respect to the RNGH instability the Benjamin--Feir instability has a much lower threshold, in agreement with the experimental evidences.\\
\begin{figure}[h!tb]
\begin{center}
\subfigure[]{\includegraphics[width=0.48\textwidth]{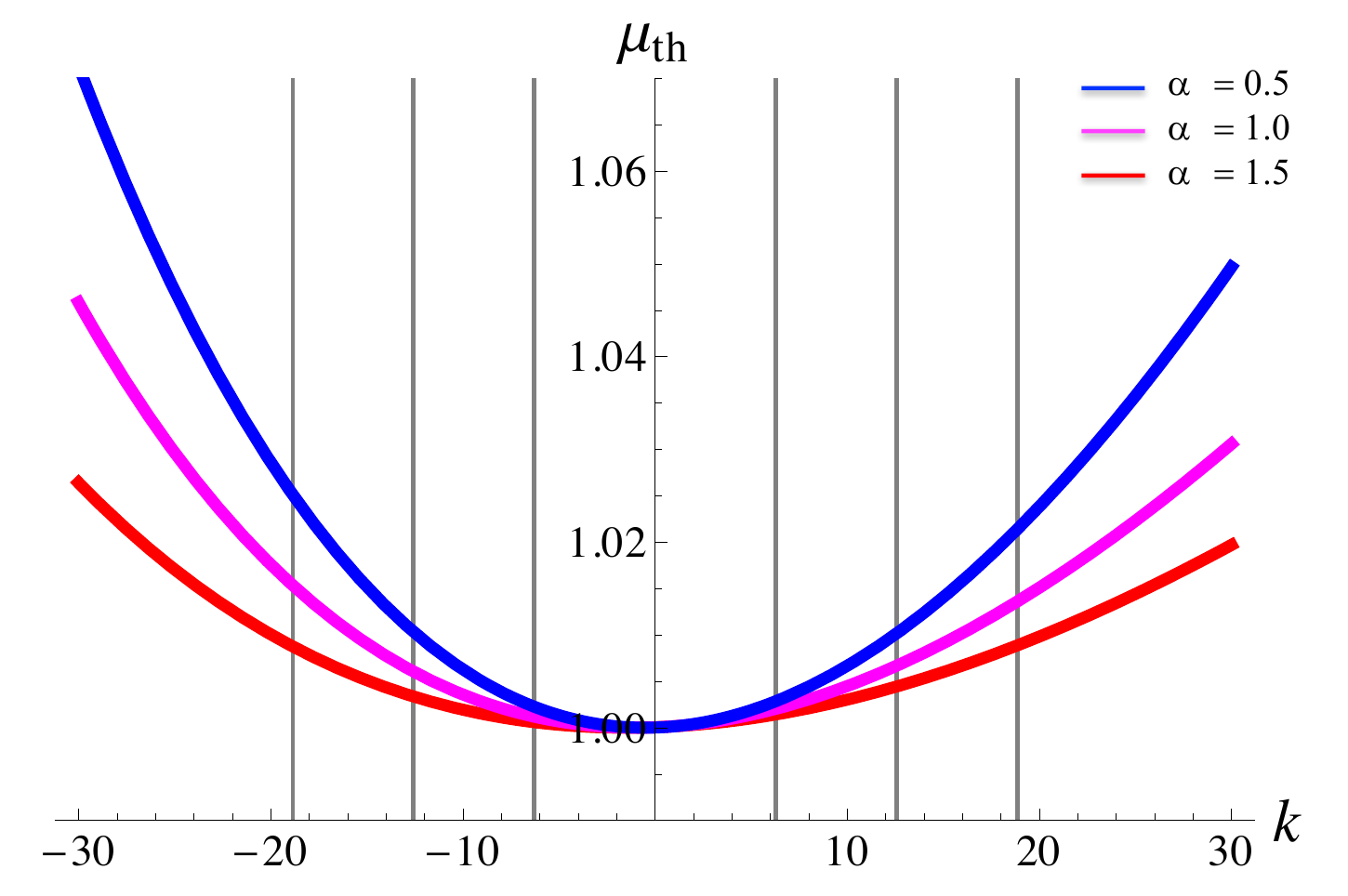} }  \hspace{0.2cm}  
\subfigure[]{\includegraphics[width=0.48\textwidth]{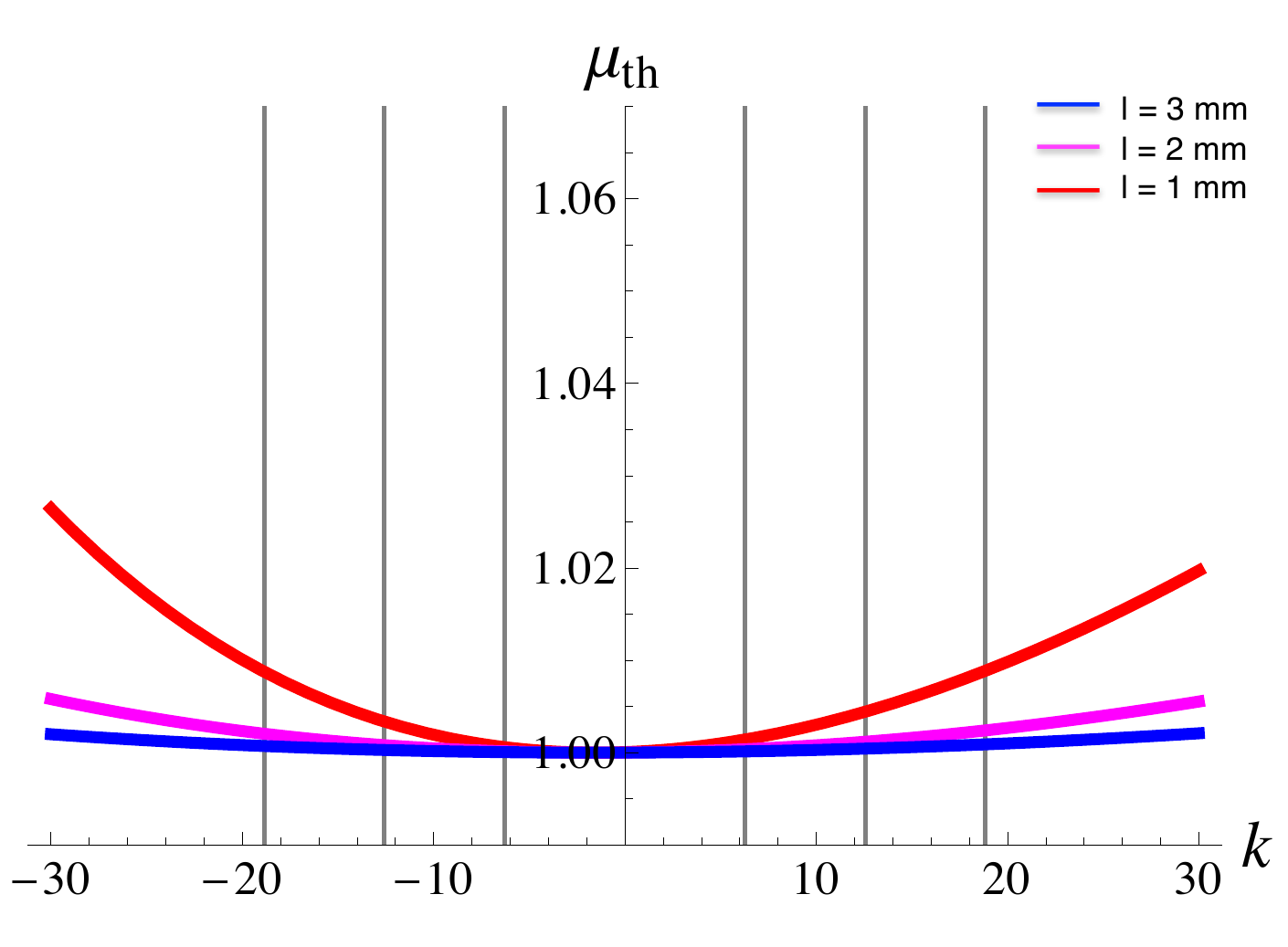}}
\subfigure[]{\includegraphics[width=0.55\textwidth]{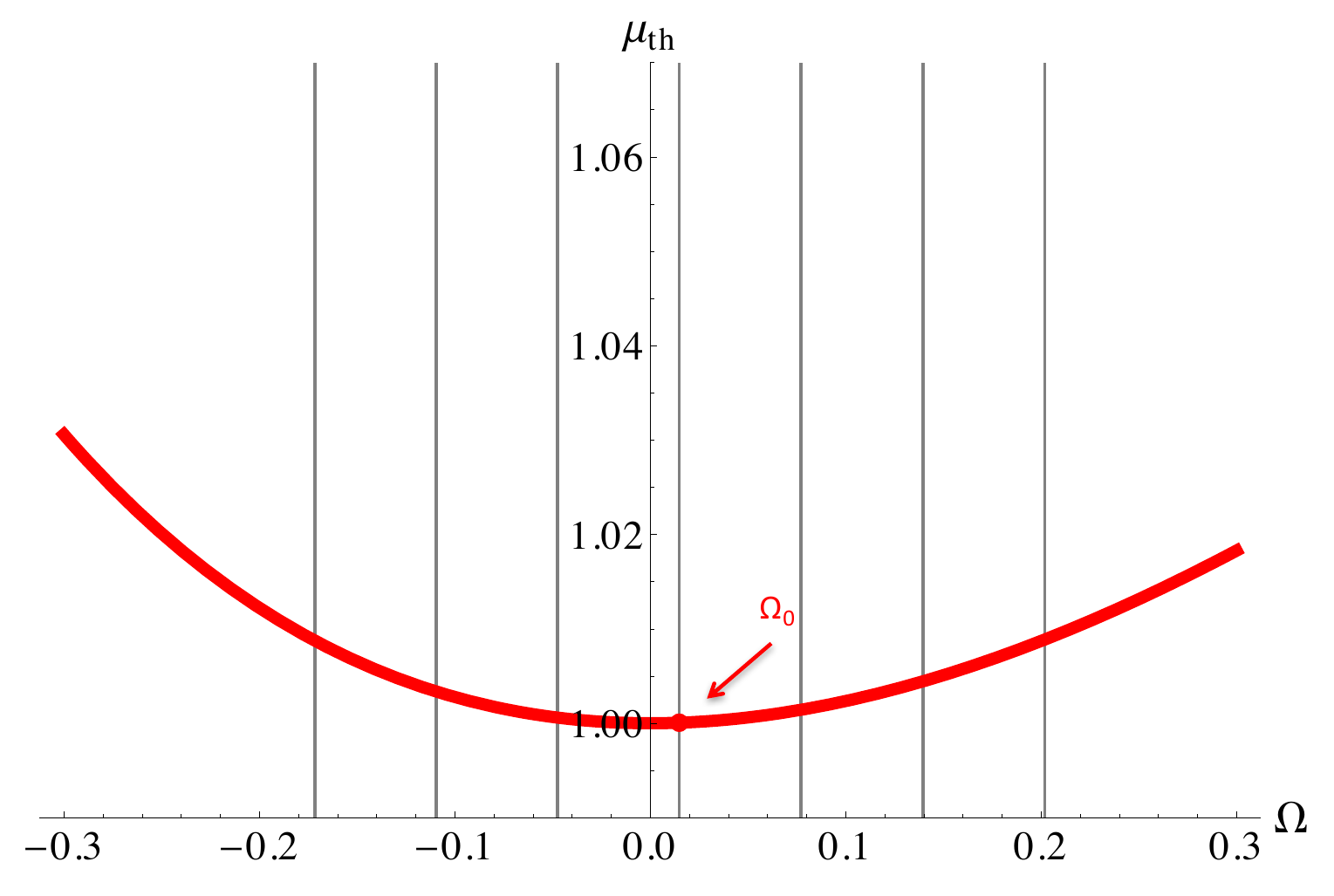}}
\end{center}
\vspace{-0.2cm}
\caption{The laser threshold plotted versus $k$ (for $\Gamma$$=$$1.1$) for different values of $\alpha$ (a) and cavity length $l$ (b). Since $z$ is scaled on $l$ the set of discrete $k$ values compatible with the boundary conditions (empty cavity modes) are separated by $2\pi$. They are indicated by continuous vertical lines in the figure. As expected, the mode $TW_{0}$ corresponds to $k=0$ and to a value of $\Omega$ given by the dispersion relation Eq. (\ref{freq}). (c) Laser threshold plotted as a function of $\Omega$ for $\alpha=1.5$, $l=1mm$ (FSR of $100 GHz$).}
\label{CWmuth}
\end{figure}
Now, in order to determine the conditions where an easier destabilization of the single mode TW$_{0}$ emitted just above the lasing threshold, or a stronger modal competition emerging therefrom, can occur we proceed to derive the analytical pump threshold curve $\mu_{th}$ of the TW lasing solutions (setting $|E_{s}|=0$ in Eq.(\ref{field})) and study it for different values of the LEF $\alpha$ and cavity length $l$ (i.e. of the FSR). As it turns out, the curve shape and the mode spectrum can provide us with a first insight about the stability of the TW$_{0}$. \\
In Fig. \ref{CWmuth}.a and Fig. \ref{CWmuth}.b we report $\mu_{th}$ as a function of $k$, considered as a continuous variable, for different values of $\alpha$ and $l$. We set $\Gamma=1.1$ that corresponds to a gain linewidth of $\simeq 1.75 THz$ which is close to that of the broad-band QCL studied in \cite{Scalari1,barbieri}.
Clearly, an increment of $\alpha$ or $l$ lowers the lasing threshold for an increasing number of TWs. An easier destabilization of $TW_{0}$ and a more complex multimode dynamics can thus be expected. In Fig. \ref{CWmuth}.c we plot the pump threshold $\mu_{th}$ against $\Omega$ for $\alpha=1.5$ \cite{green, ravaro} and $l=1mm$. As indicated by the arrow, in this case the $TW_{0}$ angular frequency is $\Omega_{0}=0.015$ that corresponds in physical units to a frequency separation of $24 GHz$ from the gain peak. This will be the case study for the following sections.

\section{Numerical results. Coherent multimode dynamics and chaotic regimes.}\label{sec3}

In this section we study the laser behavior emerging from the destabilization of the TW0 by numerical integration of the Eqs. (\ref{E4b})-(\ref{P4b1}) using a split-step method based on a second order Runge-Kutta and a FFTW algorithm.\\
In the case of Fig. \ref{CWmuth}.c, we simulated the laser dynamics for increasing values of the pump parameter $\mu$ and
we plotted the temporal evolution of the field intensity, the optical spectrum (OS) and the beat note spectrum (BNS) for specific values of the normalized pump parameter. 
\begin{figure}[h!]
\begin{center}
\subfigure[]{\includegraphics[width=0.55\textwidth]{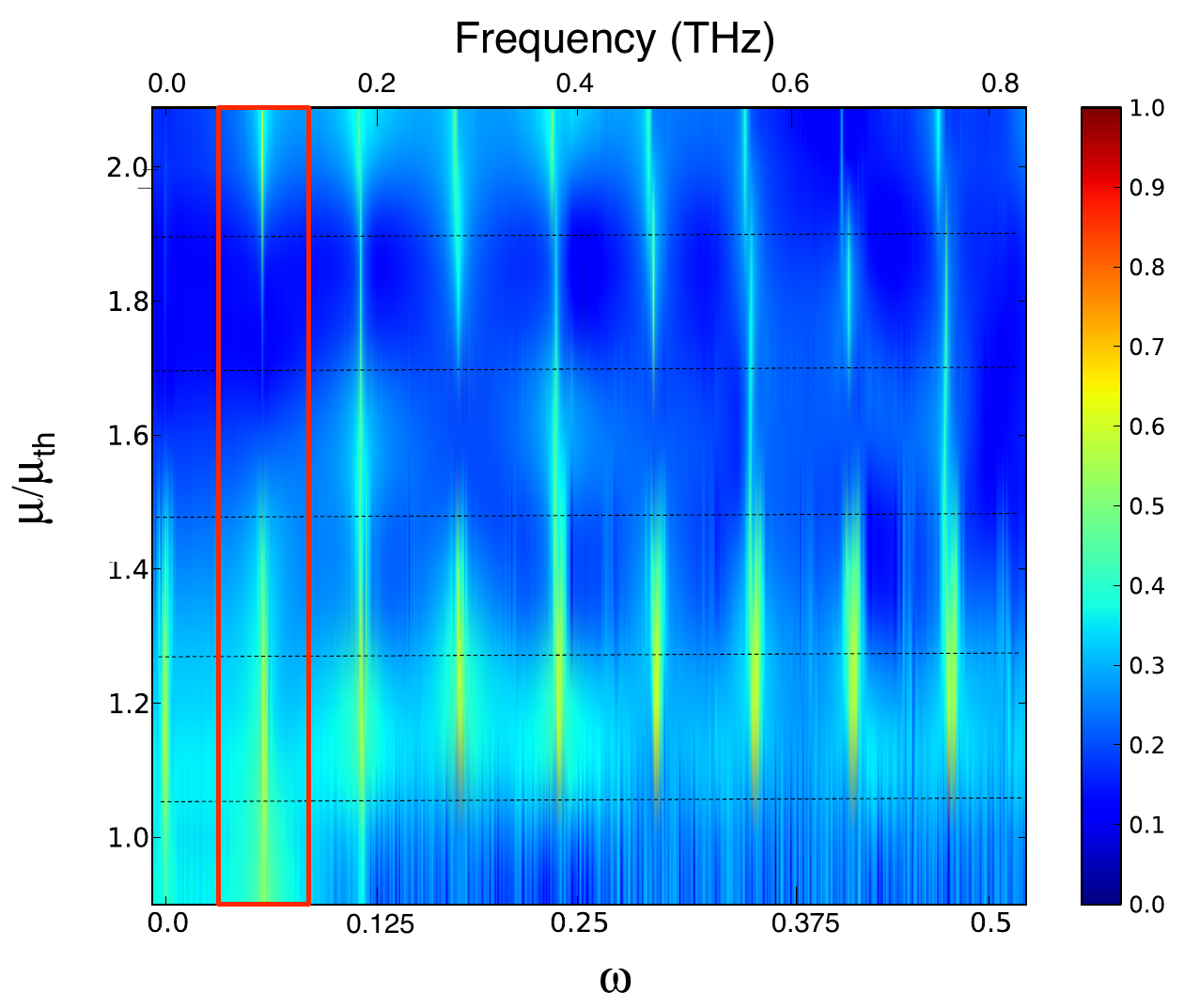} \hspace{0.3cm} \includegraphics[width=0.55\textwidth]{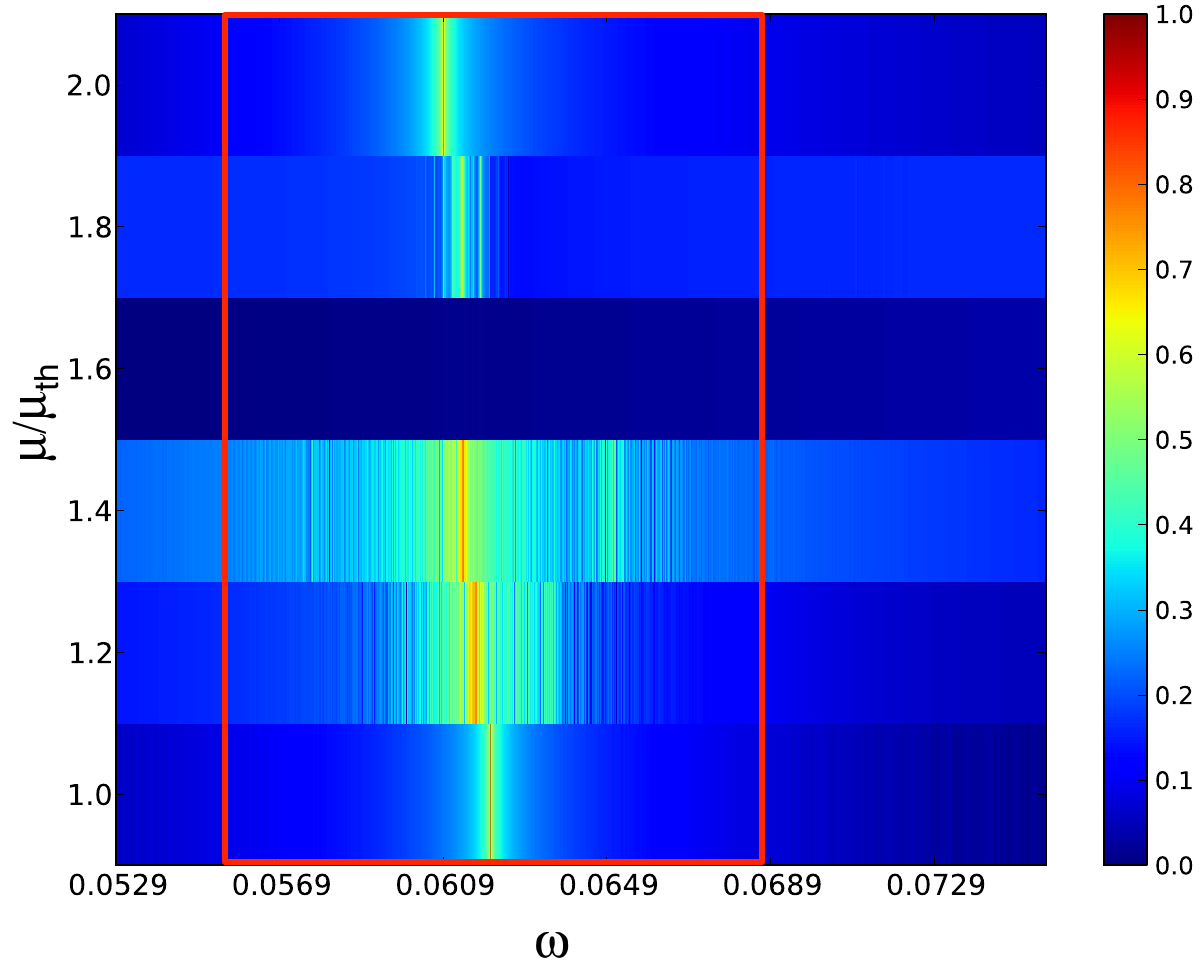} } \\
\subfigure[]{\includegraphics[width=0.57\textwidth]{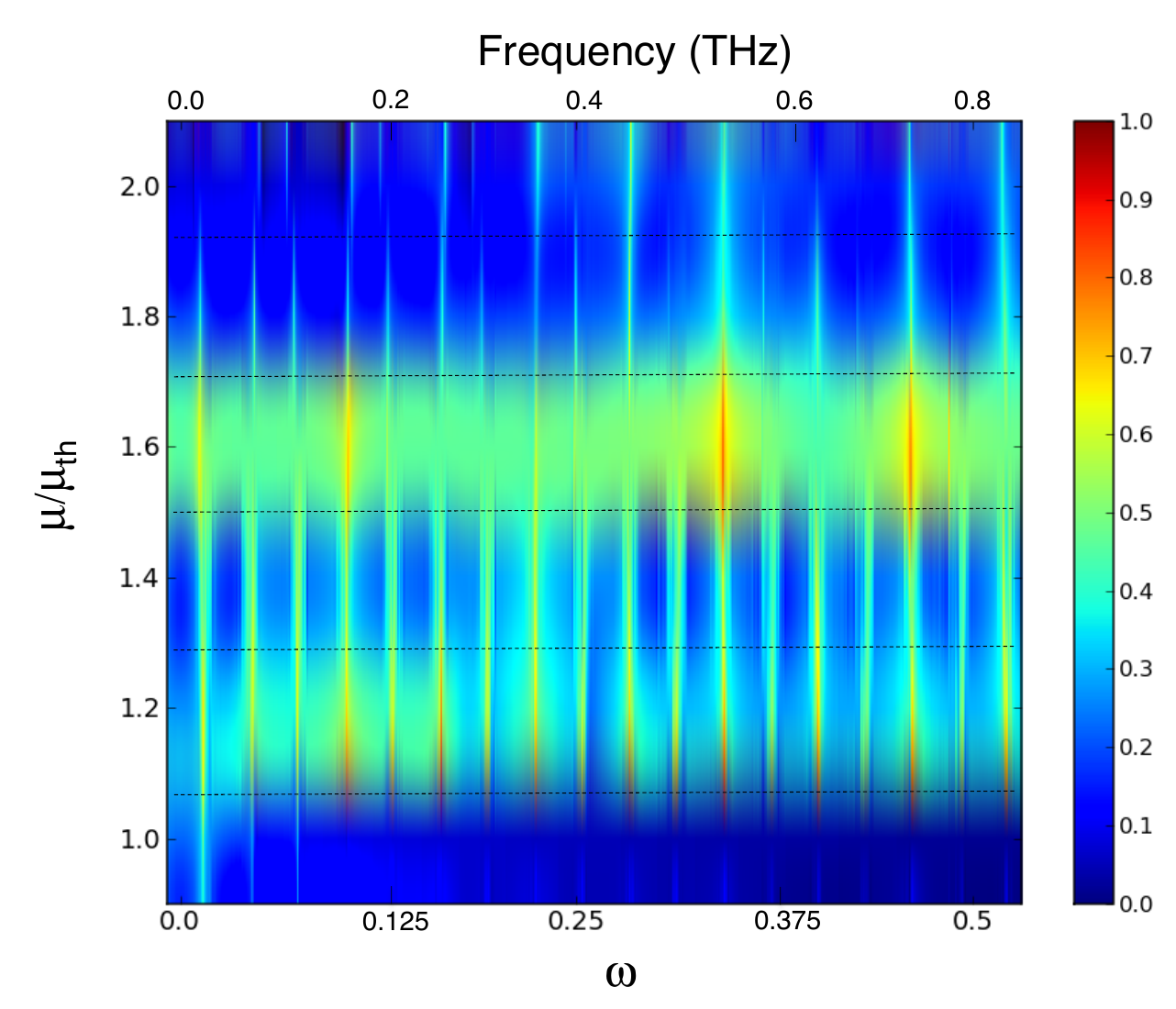}}
\end{center}
\vspace{-0.2cm}
\caption{Beat note spectrum ((a), left panel) and its zoom on the first beatnote ((a), right panel) and optical spectrum (b) of the QCL field for different values of the normalized pump parameter. For an easier comparison with experimental data, the frequency in physical units is drawn on the upper horizontal scale. The color scale is logarithmic and covers about $16$ orders of magnitude. All spectra have been shifted to have the same minimum, taken as zero and the maximum corresponds to the absolute maximum among all spectra. During the whole simulation $\mu$ is adiabatically increased by steps of $0.2$ leaving the dynamical system described by Eqs.(\ref{E4b})-(\ref{P4b1}) to reach a regime. The  dashed lines delimit the regions where the value of $\mu$ is kept constant. Color scale variations across these regions are due to graphical interpolation.
The red rectangle highlights the region of the BNS where analogous experimental data are available \cite{barbieri}. Parameters as in Fig. \ref{CWmuth}.c.}
\label{beatnote}
\end{figure}
In Fig. \ref{beatnote}.a, just above threshold ($\mu = 1.002 \mu_{th}$ with $\mu_{th}=\mu_{th}(k=0)=1$), we find as expected the emission of $TW_{0}$, affected by a self--oscillation, whose signature is seen in the BNS peak at $0.062$ (quite close to the normalized FSR $=(2 \pi c/nl)\tau_{d}=0.0628$ that corresponds to a frequency separation of $100 GHz$ in physical units) and the presence of higher angular frequencies separated by $\simeq$ FSR. Consistently, in the OS of Fig. \ref{beatnote}b, we observe the main peak at the emission angular frequency $\Omega_{0}=0.015$ of the solution $TW_{0}$, according to Eq. \ref{freq}, weaker peaks at $\simeq \Omega_{0}+FSR$ and $FSR-\Omega_{0}$ that are the signature of nonlinear parametric processes. The corresponding dynamical regime consists of regular small amplitude oscillations around an almost constant intensity.\\
Beyond the immediate proximity of the threshold ($\mu$ $>$ $1.002 \mu_{th}$), a multimode instability takes place and brings the system in a turbulent spatio-temporal regime. An example of the temporal plot of the field intensity taken at $z=1/2$ is shown in Fig. \ref{dyn1} for $\mu=1.2 \mu_{th}$. The BNS of Fig. \ref{beatnote}.a shows the broadening of the first BN and the appearance of higher beatnotes separated by $\simeq$ FSR. The corresponding OS of Fig. \ref{beatnote}.b shows the same, but the broadening of the QCL modes is more pronounced and the wealth of modes setting off is more evident.
\begin{figure}[h!t]
\begin{center}
\subfigure[]{\includegraphics[width=0.48\textwidth]{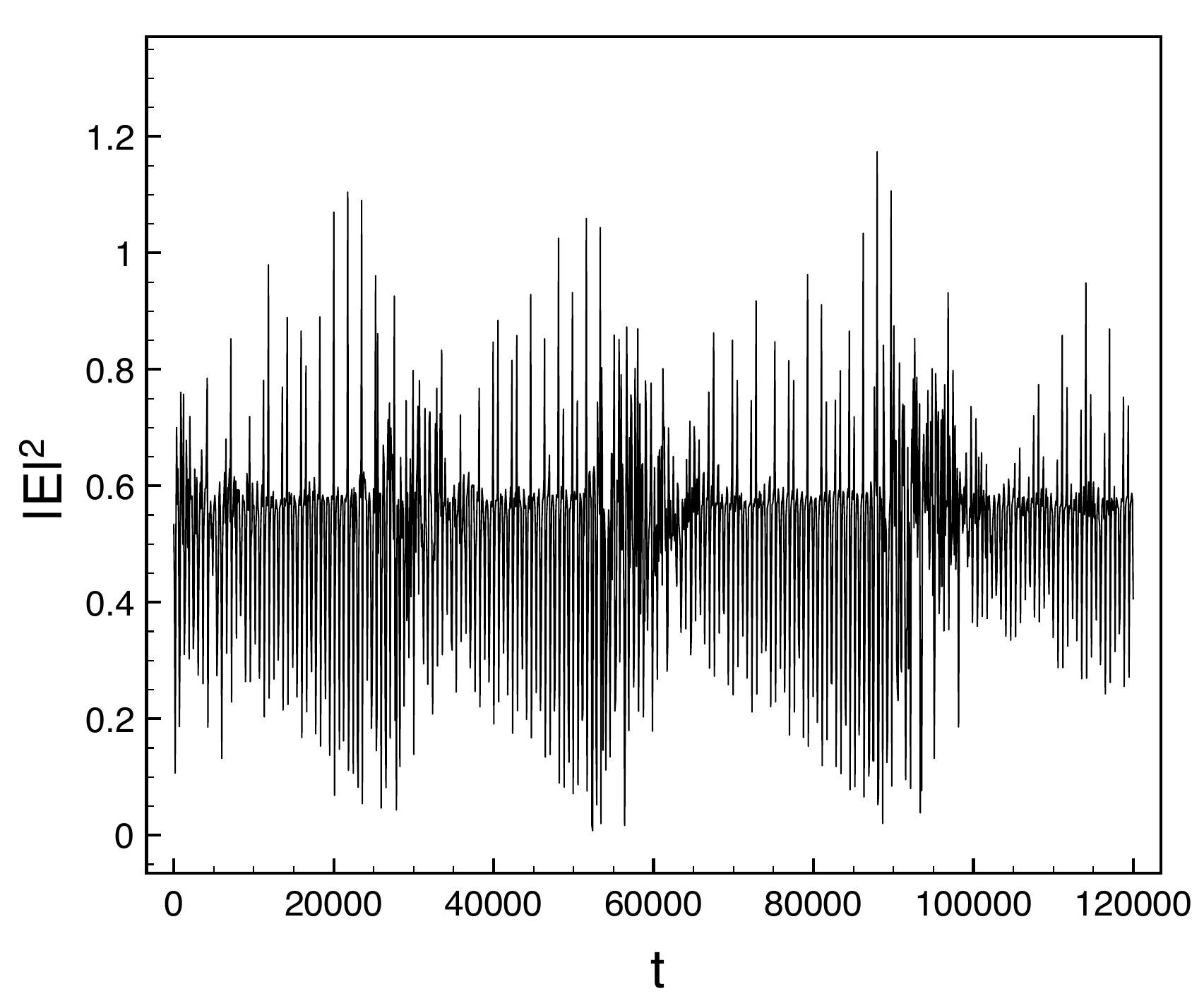}}
\subfigure[]{\includegraphics[width=0.48\textwidth]{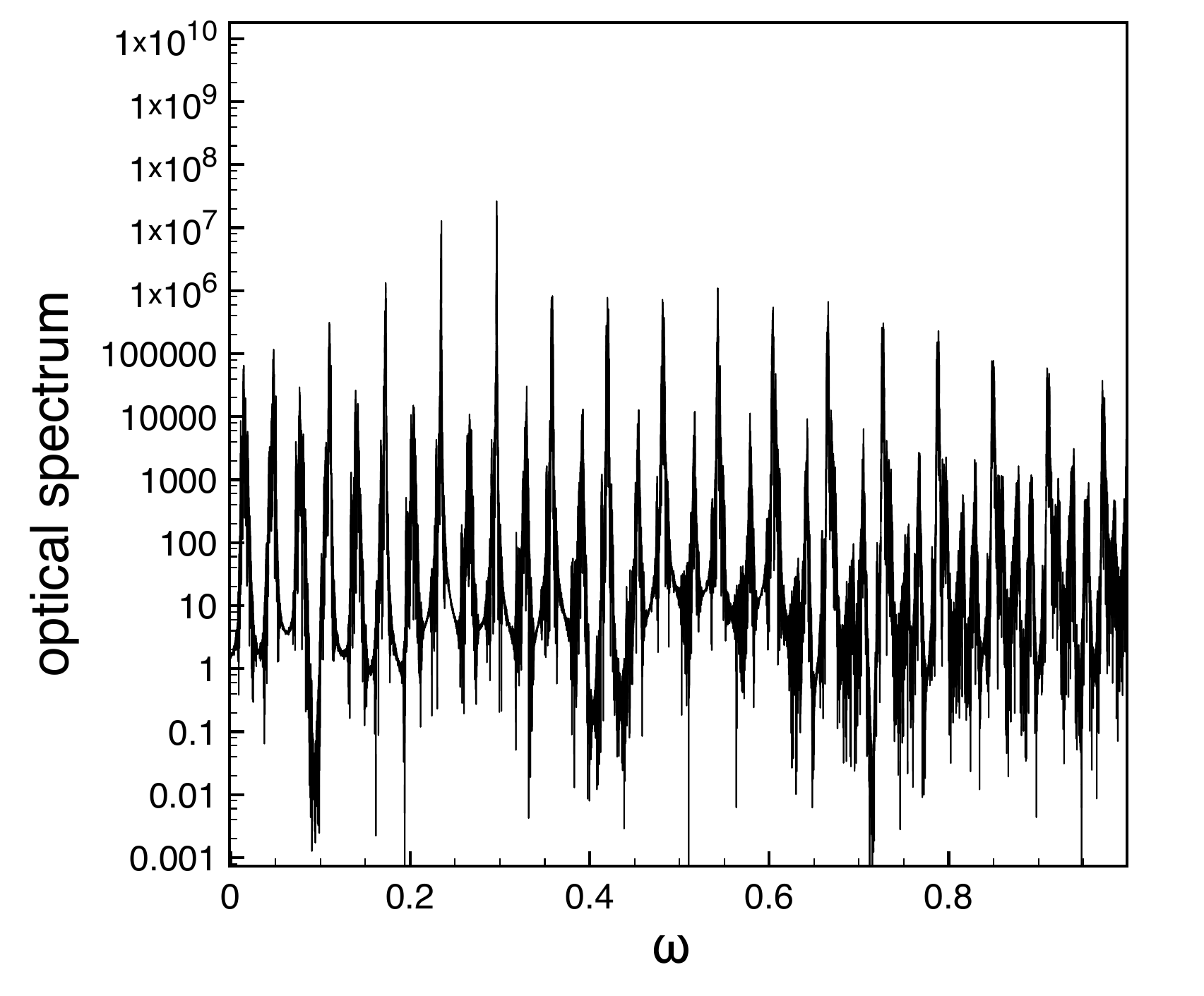}}
\end{center}
\vspace{-0.2cm}
\caption{Intensity at half the cavity versus time (a) and corresponding OS (b) in the irregular multimode regime corresponding to $\mu=1.2 \mu_{th}$. Time is scaled to $\tau_{d}=0.1ps$. Other parameters as in Fig. \ref{beatnote}.}
\label{dyn1}
\end{figure}
\noindent A similar irregular behavior is found for $\mu=1.4 \mu_{th}$. The next value of $\mu=1.6 \mu_{th}$ brings the system back to a regular regime whose field intensity dynamics is shown in Fig. \ref{14spectr}, along with the OS. At steady state, both the plot of the intracavity field intensity at $z=1/2$ versus time and the longitudinal intensity profile along the resonator at a fixed time reveal the presence of a self-confined, two-peaked structure that indefinitely travels in the ring resonator with constant speed and shape, as shown in Fig. \ref{puls}. It represents a phenomenon of spontaneous mode-locking, although it involves a limited number of modes. The corresponding OS is in fact characterized by only $7$ active modes within two decades (see Fig. \ref{14spectr}.b). The presence of two traveling peaks is of course associated to an intensity spectrum showing a periodicity at half the round trip time, i.e. at twice the BN angular frequency. This feature can thus explain the disappearance of the fundamental BN observed in Fig. \ref{beatnote}.a and in certain dynamical regimes reported in \cite{barbieri} (see for example Fig.4 of \cite{barbieri} in the region between $400$ and $420$ mA). We could \textit{a posteriori} verify that this traveling structure is also present in the field profile just above threshold ($\mu=1.002 \mu_{th}$) where, of course, the low absolute emission leaves a very low-contrast. We also observe that these results are stable upon inclusion of an additive, stochastic noise process in the simulations.

\begin{figure}[h!t]
\begin{center}
\subfigure[]{\includegraphics[width=0.48\textwidth]{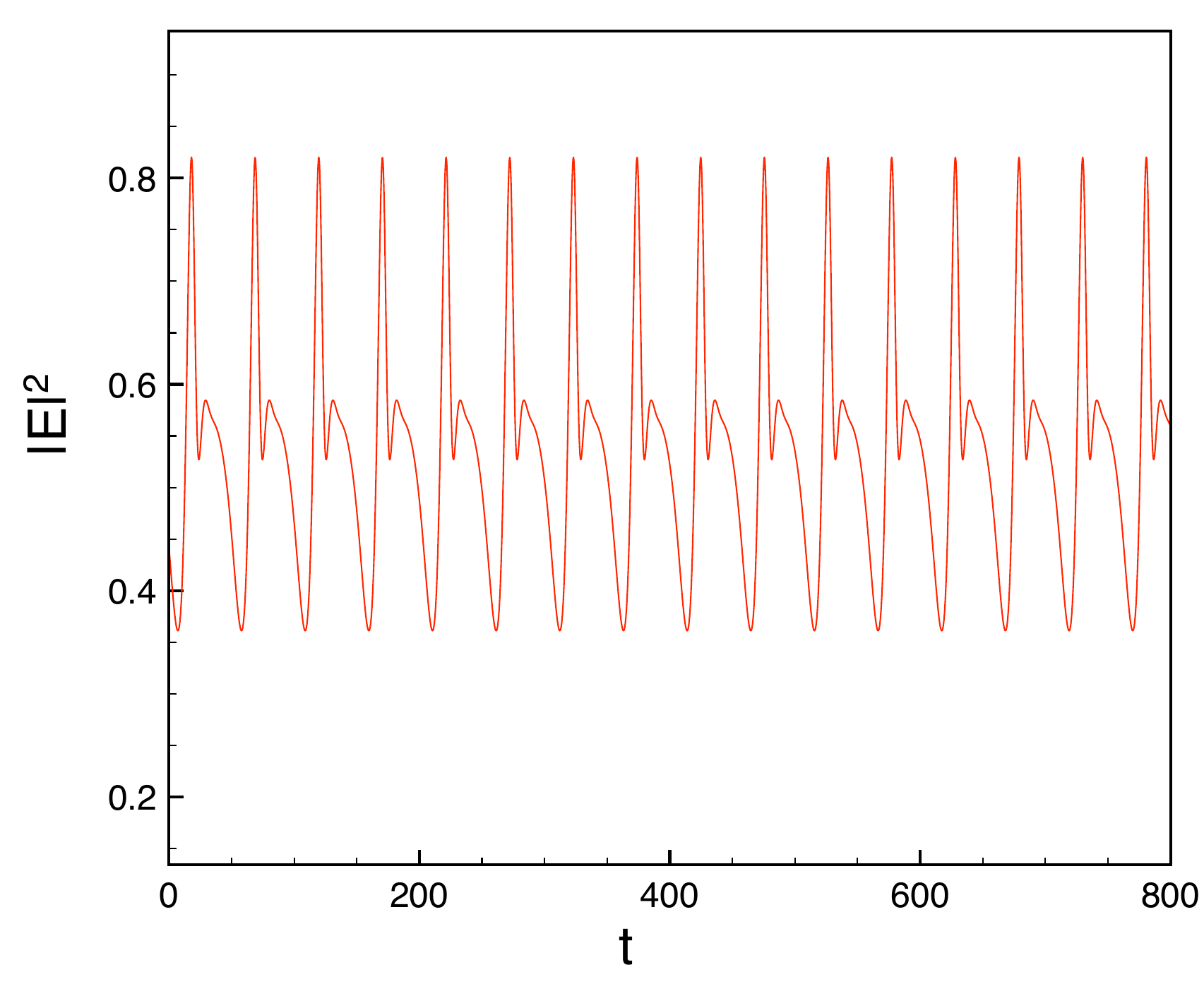}}
\subfigure[]{\includegraphics[width=0.48\textwidth]{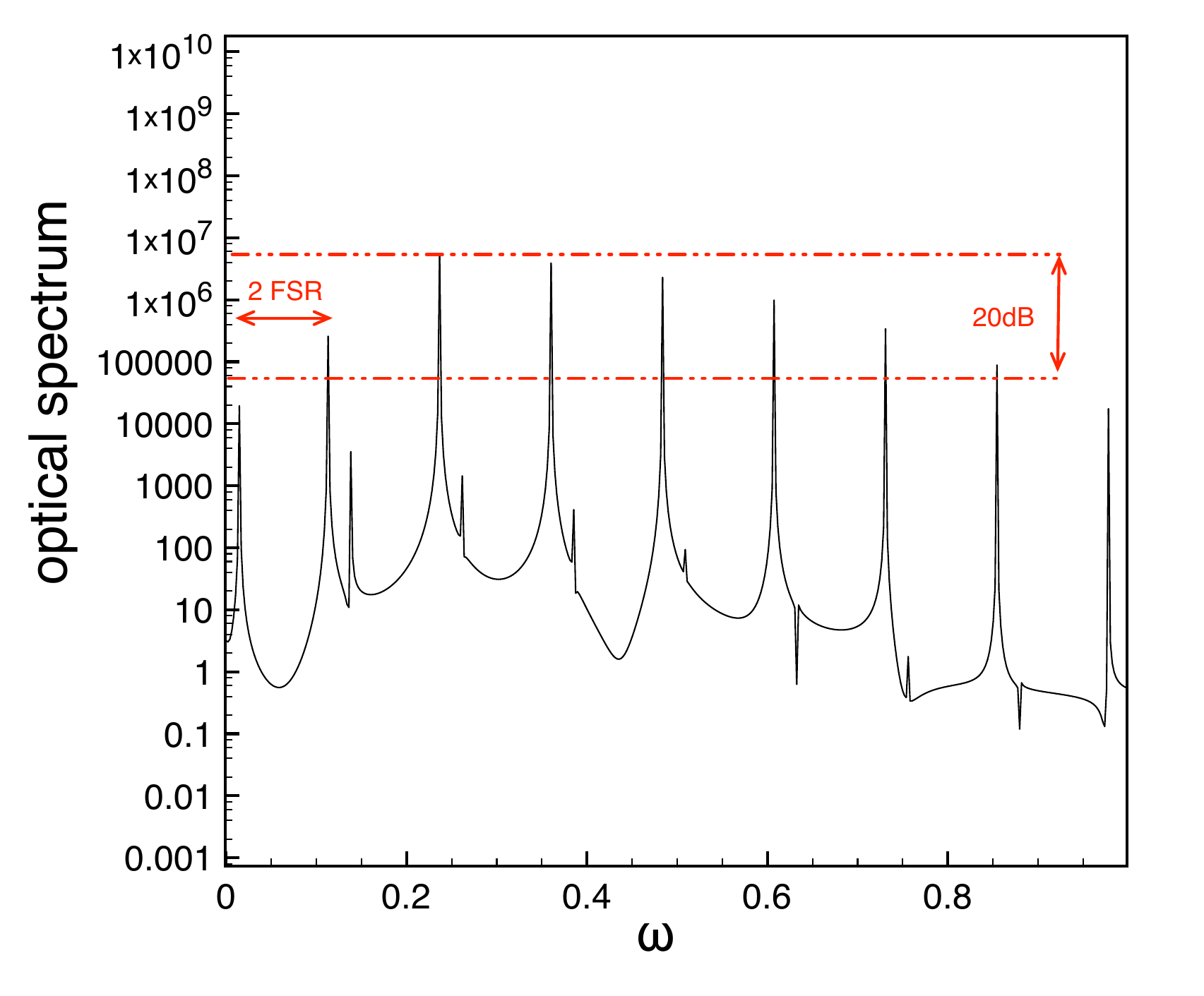} }
\end{center}
\vspace{-0.2cm}
\caption{Intensity at half the cavity versus time (a) and corresponding OS (b) in the regular multimode regime for $\mu=1.6 \mu_{th}$. Note the main peaks at integer multiples of $\simeq 2\times FSR$. Other parameters as in Fig. \ref{beatnote}.}
\label{14spectr}
\end{figure}
\begin{figure}[h!t]
\begin{center}
\includegraphics[width=0.55\textwidth]{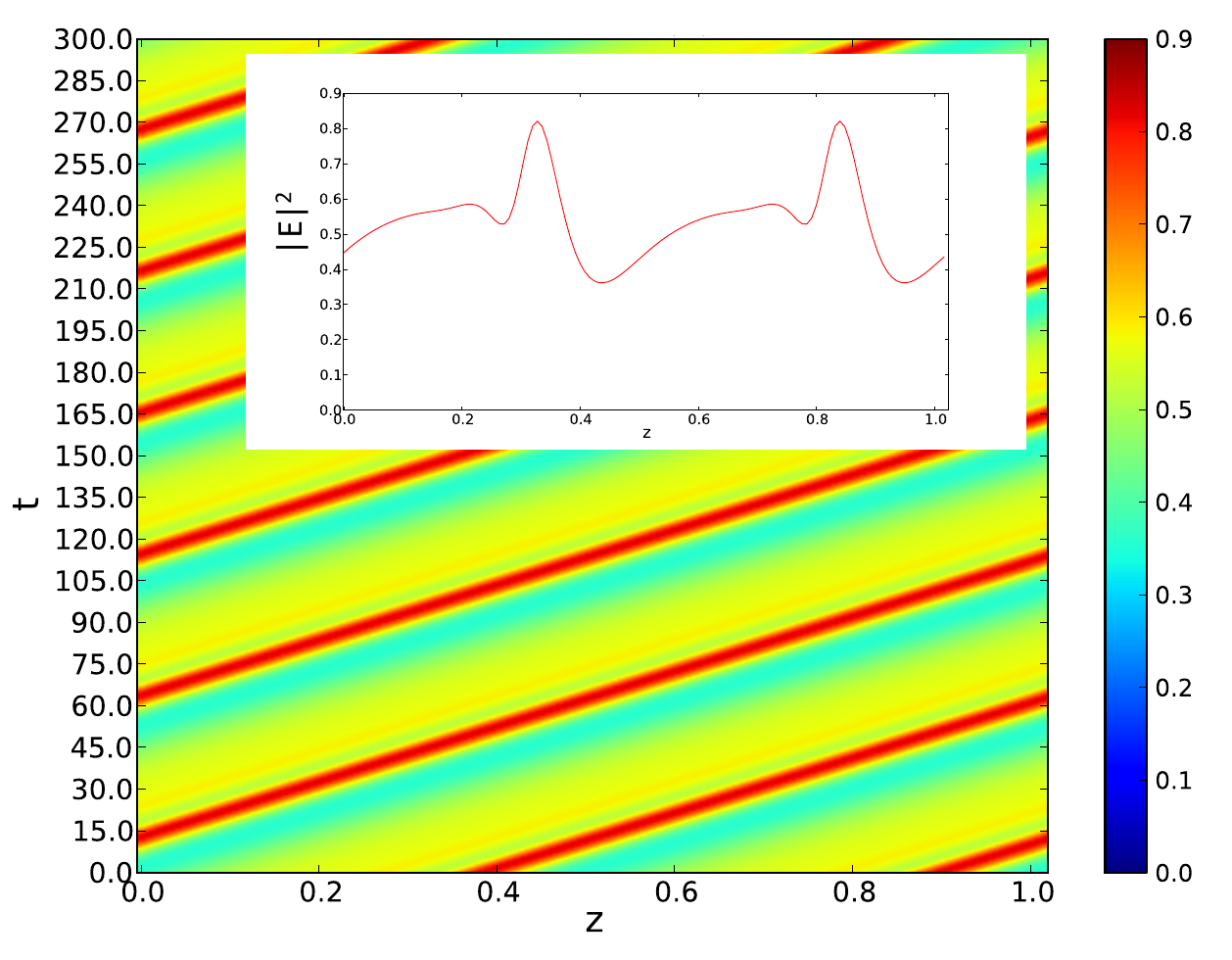}
\end{center}
\vspace{-0.2cm}
\caption{Spatio-temporal evolution ($z$:cavity coordinate, $t$:time scaled on $\tau_{d}$) of the intracavity intensity and (inset) spatial intensity profile at a given instant for $\mu=1.6 \mu_{th}$. Other parameters as in Fig. \ref{beatnote}. }
\label{puls}
\end{figure}
\begin{figure}[h!t]
\begin{center}
\includegraphics[width=0.7\textwidth]{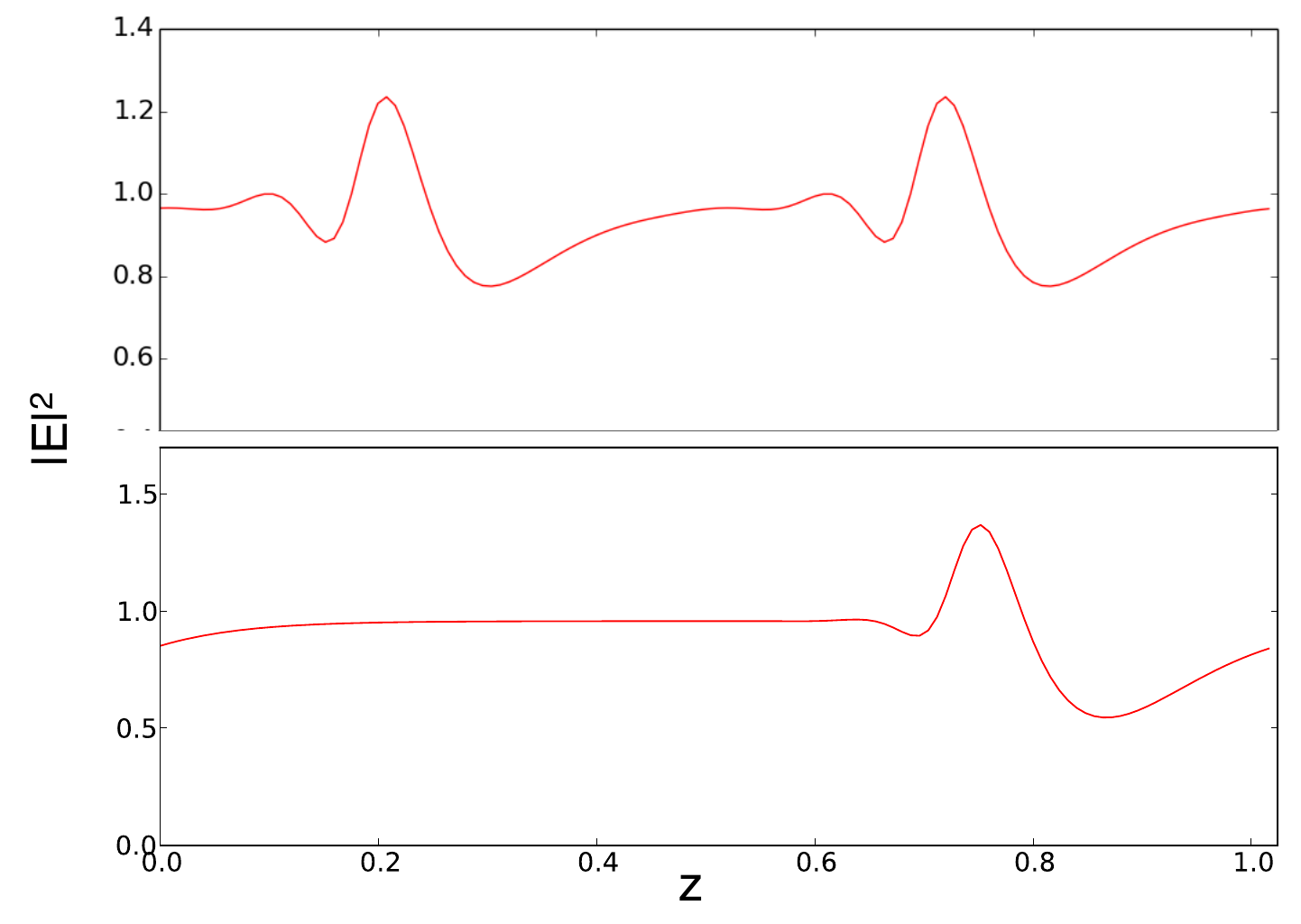}
\end{center}
\vspace{-0.2cm}
\caption{Intensity profile at a given instant of time obtained starting from two different initial conditions at $\mu=2.0 \mu_{th}$: a single peak structure (down) and a two-peaked structure (up). Other parameters as in Fig. \ref{beatnote}. }
\label{puls2}
\end{figure}

\noindent
%
At $\mu=1.8 \mu_{th}$ and at $\mu=2.0 \mu_{th}$ the dynamics is still regular in both cases (the number of locked modes in the OS in the two regimes is $9$ and $10$ within two decades, respectively).
A proof in support of the solitonic character of the travelling structures described above is provided at $\mu=2.0 \mu_{th}$. Figure. \ref{puls2} shows how two different initial conditions can evolve at steady state into a single--peaked self-localized structure or to a two-peaked one propagating at a constant speed, a phenomenon of multistability, typical of dissipative solitons in extended systems \cite{Akhmediev}. Up to $N=3$ stable structures could be obtained at regime from different initial conditions.\\
%

\begin{figure}[h!t]
\begin{center}
\includegraphics[width=0.75\textwidth]{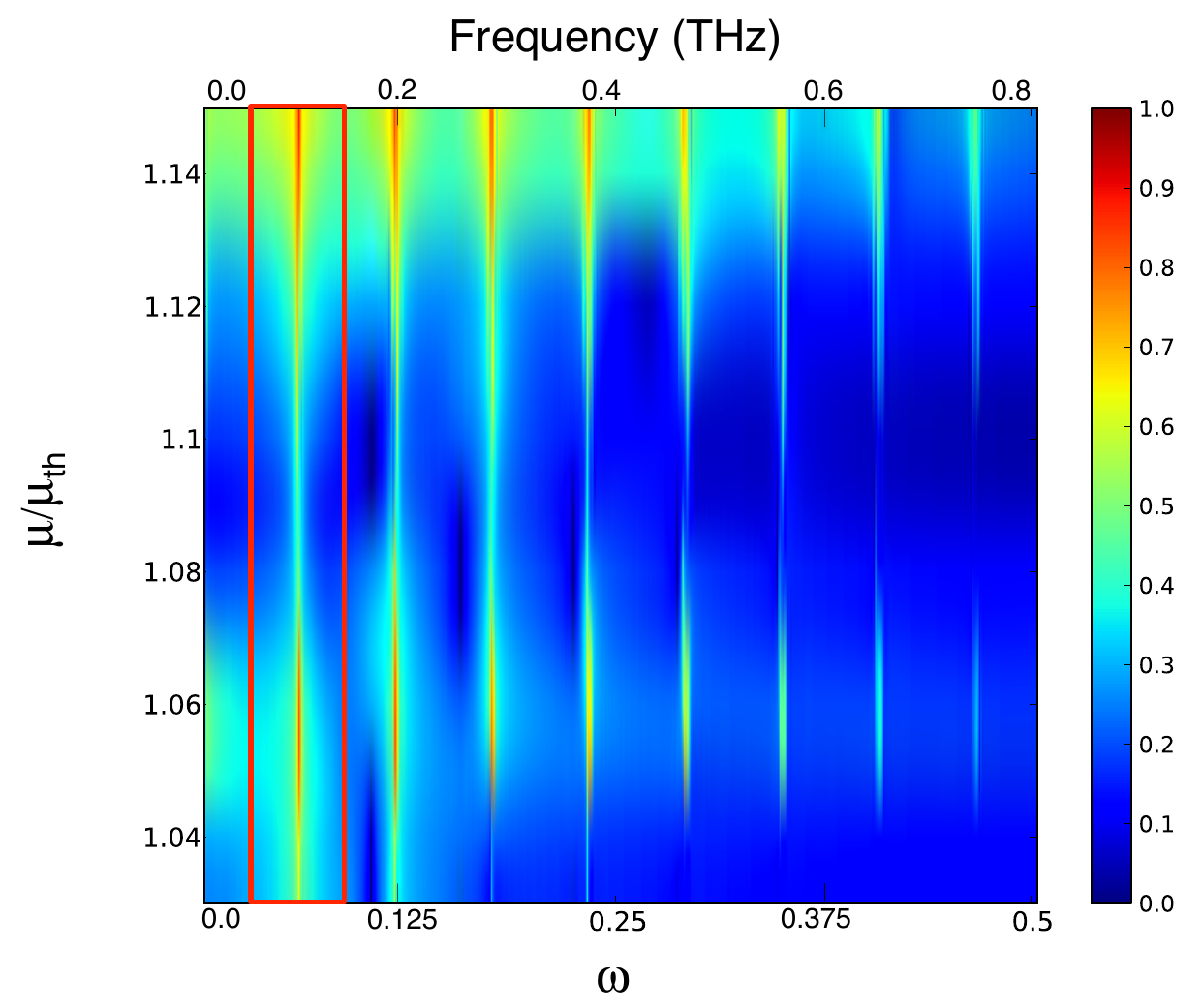}
\end{center}
\vspace{-0.2cm}
\caption{$\Gamma=0.3$. Other parameters as in Fig. \ref{beatnote}. BNS of the emitted field above threshold for increasing$\mu/\mu_{th}$. Same color scale as in Fig. \ref{beatnote}. $\mu$ is adiabatically increased by $0.02$.}
\label{03spectra}
\end{figure}

%
We may thus conclude that a good qualitative agreement exists with the results reported e.g. in Fig. 4 of \cite{barbieri}, as our model properly reproduces (though for different values of the the bias-to-threshold current ratio) a sequence of narrow lines at multiples of the FSR, and broad bands. \\
The absence of the multibeatnote regimes reported in \cite{barbieri} (see Fig. 5 in \cite{barbieri} at pump $440$ mA) can be explained by the homogeneous gain line assumed by our model, as opposed to the inhomogeneous line of the THz-QCL in \cite{barbieri} which integrates three different active regions in the same waveguide. The different gain peaks probably introduce different BN with the common resonator. \\
Moreover, our numerical simulations show that the extension of the regions of regular and irregular system dynamics, as well as the number of competing longitudinal modes can be controlled by acting on the QCL intrinsic parameters such the LEF $\alpha$, the carrier decay time $\tau_{e}$, the cavity length $l$ and the FWHM of the gain curve ($\sim \Gamma$).
As an example, in Fig. \ref{03spectra} we report the BNS for a value of $\Gamma$ almost three times smaller than that used to produce Fig. \ref{beatnote} and hence comparable with the gain linewidth of the single stack active region considered in \cite{Scalari1,barbieri}. Although, as expected, the number of competing modes becomes smaller, we observe a dynamical scenario characterized by a stable TW emission up to $\mu=1.04 \mu_{th}$, and alternating windows of regular oscillations and chaotic dynamics for $1.04<\mu/\mu_{th}<1.14$.
%

\subsection{Effect of RF modulation}

\begin{figure}[h!t]
\begin{center}
\includegraphics[width=0.55\textwidth]{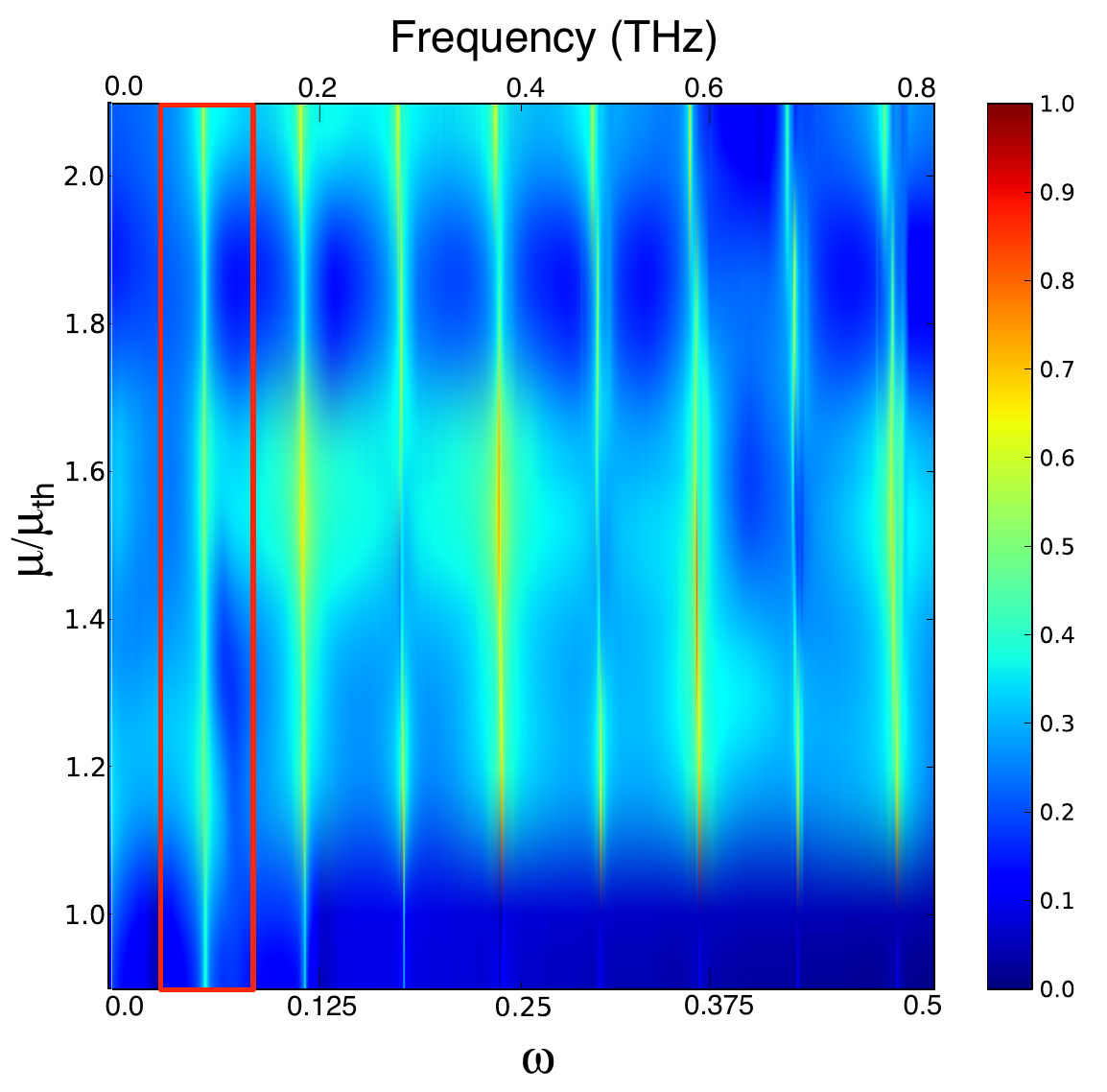}
\end{center}
\caption{Beat note spectrum of the electric field emitted by the QCL in presence of a RF modulation of amplitude $10\%$ applied to $2/5$ of the QCL at a angular frequency of $\Omega_{M}=0.062$. The BN, as evidenced by the red box, is now present throughout the whole spectrum. Parameters as in Fig. \ref{beatnote}.}
\label{mod}
\end{figure}

\begin{figure}[h!t]
\begin{center}
\includegraphics[width=0.55\textwidth]{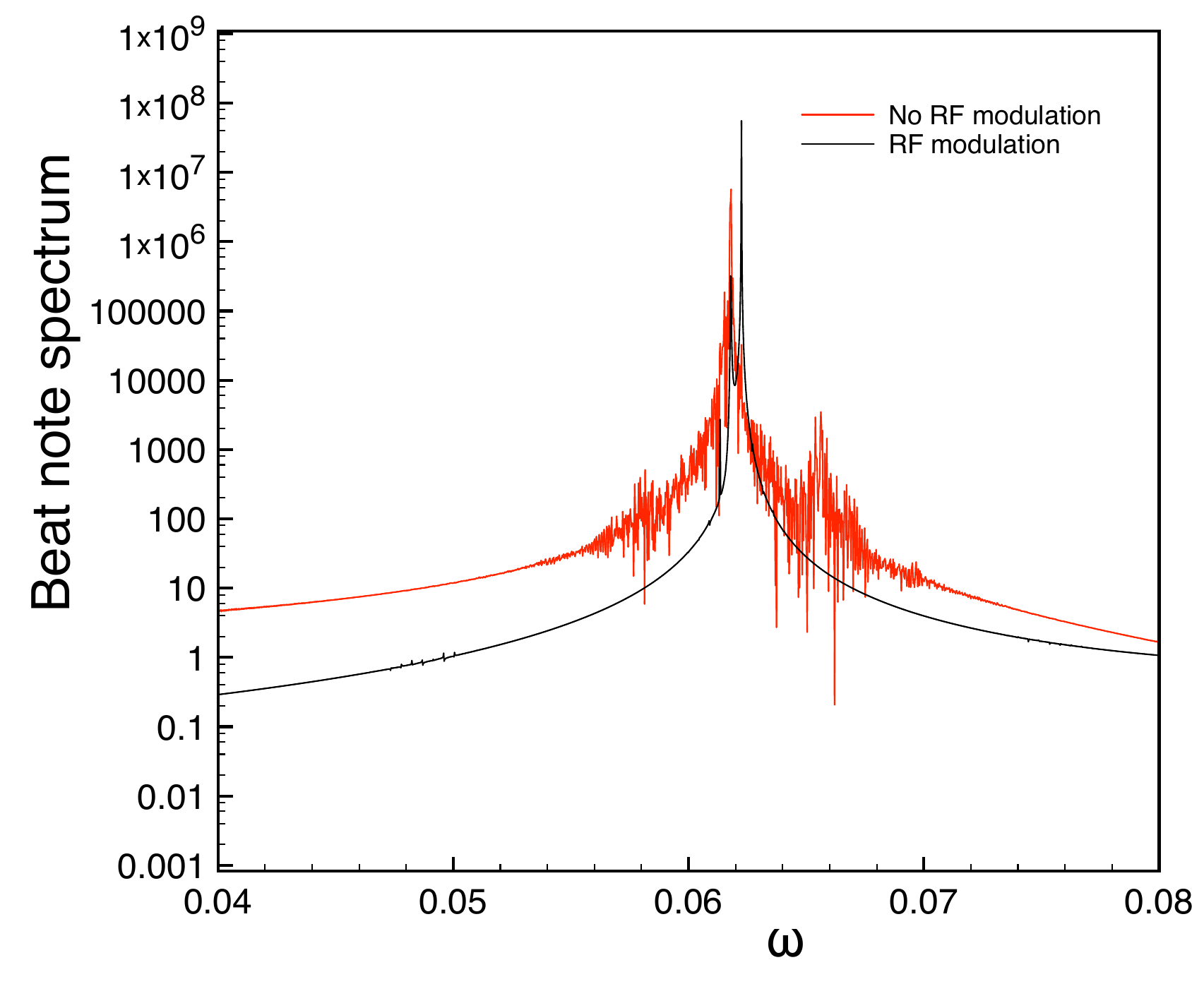}
\end{center}
\caption{Beat note spectrum of the electric field emitted by the QCL in presence (black line) and absence (red line) of RF pump modulation. In the simulations we used $\mu_{0}=1.4 \mu_{th}$ and a RF modulation, applied only to a just $2/5$ of the whole laser cavity, with an amplitude equal to $35\%$ of the mean value and $\Omega_{M}=0.062$. Other parameters as in Fig. \ref{beatnote}. }
\label{modplot}
\end{figure}

An interesting result reported in \cite{barbieri} was the line narrowing in the BNS caused by the application of a RF modulation to the current pump. The narrowing was interpreted as the onset of comb coherence cased by the external "forcing". \\
Our model proved capable to reproduce this effect, though only qualitatively, when an external RF modulation is added to the pump current to just a small portion of the cavity in the form $\mu(t)=\mu_{M}(t)=\mu_{0}+\mu_{A}cos(\Omega_{M}t)$, with $\mu_{A}=0.1\mu_{0}$ and where $\Omega_{M}$ is close to the first beatnote observed in absence of modulation.
In the remaining part of the cavity a DC bias a $\mu=\mu_{0}$ was maintained.
As shown in Fig. \ref{mod}, the first evidence is that the addition of the RF modulation causes the fundamental BN to reappear for pump values where it was absent (in agreement with experimental evidences, as in Fig. 10 of \cite{barbieri}).
The second effect, better illustrated in Fig. \ref{modplot}, is the important reduction of the BN linewidth with a transition to coherent dynamics. The phase-noise reduction is in fact a clear manifestation on an increased coherence in the multimode emission. Starting from a chaotic state at $\mu_{0} =1.4\mu_{th}$  (whose spectrum is the red line in Fig.\ref{modplot}) we observed a transition to a regularly oscillating (multimode) state when the RF is turned on (black line in Fig.\ref{modplot}). The qualitative transition is reproduced, tough the line shrinking is much less pronounced than that reported in experiments \cite{barbieri}. In this respect we note that, at difference from the experiment, a RF applied to the whole cavity does not yield the same effect and it leaves the system in the chaotic state with little effect on the BNS. A possible explanation of this apparent discrepancy may lie in microwaves attenuation, estimated of few dB/mm \cite{barbierimicro}, that removes spatial uniformity in the experiments.\\

\section{Active mode-locking}\label{sec4}
Finally, we checked that our model could robustly reproduce active mode--locking with the generation of picosecond pulses, as experimentally verified \cite{Barbieri0, Scalari2016, Barbieri2017, Freeman}. To this purpose we applied a strong current modulation at the round trip frequency $\omega_{M}$ to a short cavity section. In particular we set $$\mu(t)=\mu_{M}(t)=\mu_{0}+\mu_{A}cos(\Omega_{M}t)\quad \mu_{0}, \mu_{A}>\mu_{th}$$ in $1/7$ of the cavity length, while in the remaining part a standard DC bias ($\mu=\mu_{b}<\mu_{th}$) is maintained.

\begin{figure}[h!t]
\begin{center}
\subfigure[]{\includegraphics[width=0.45\textwidth]{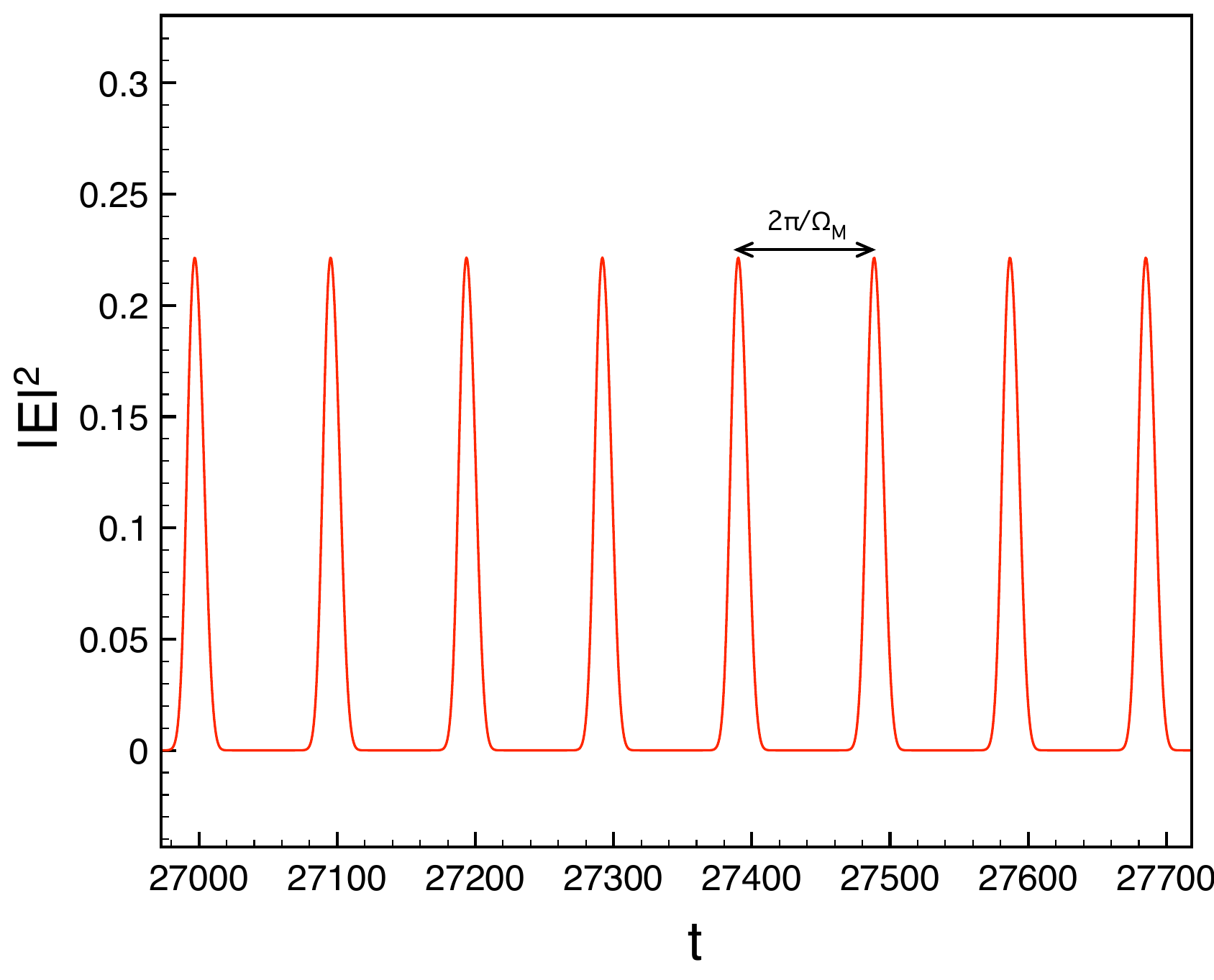}}
\subfigure[]{\includegraphics[width=0.45\textwidth]{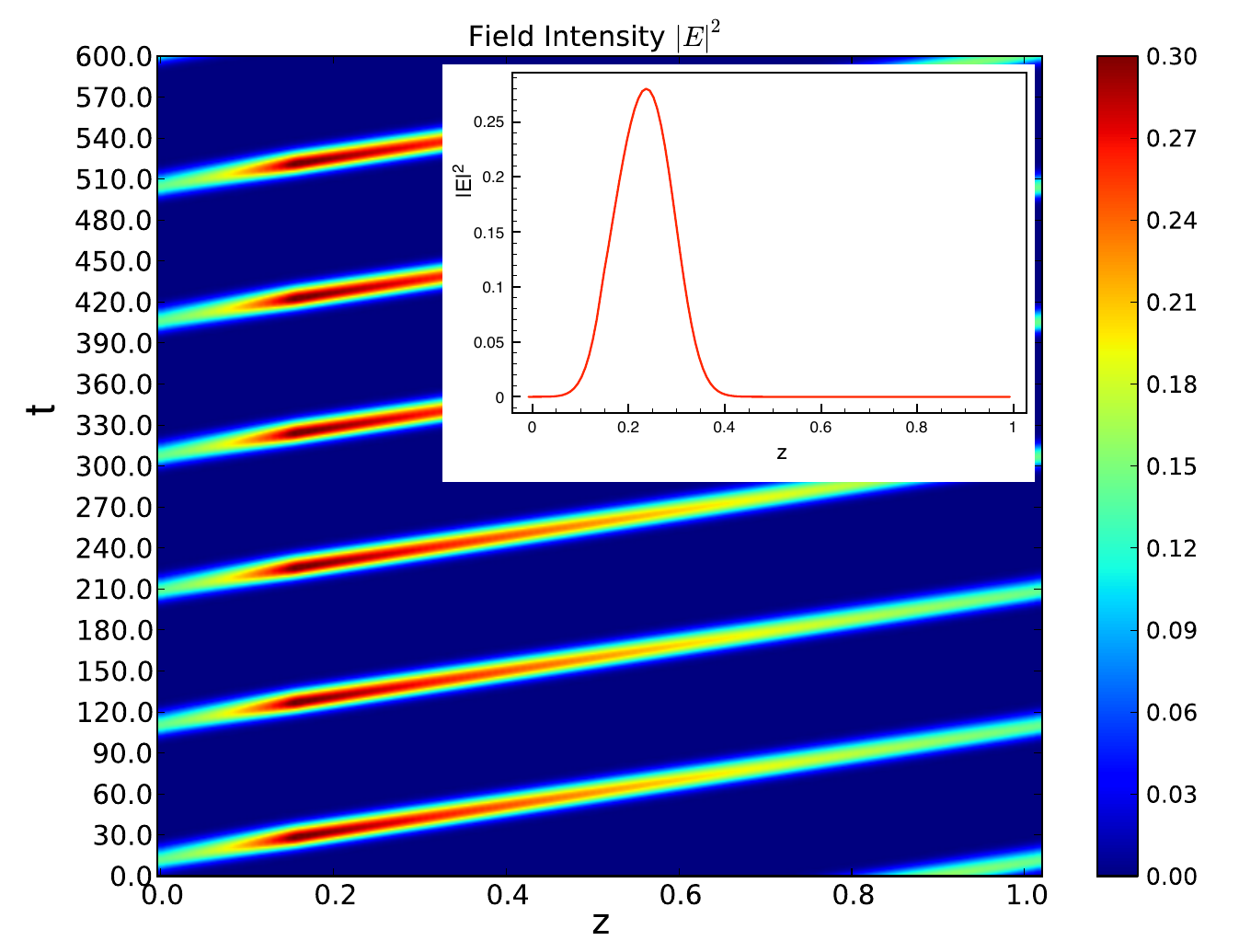}}
\end{center}
\vspace{-0.2cm}
\caption{$\alpha=1.2$. Other parameters as in Fig. \ref{beatnote}. Intensity at half the cavity (a) and intracavity intensity (b) versus time at steady showing the formation of ultrashort pulses of FWHM $\simeq 1.5$ $ps$ via active mode-locking for a sinusoidal modulation of the pump at the angular frequency $\Omega_{M}=0.0628$ ($\mu_{0}=1.5 \mu_{th}$, $\mu_{A}=4 \mu_{th}$) and a DC biased in the remaining section of the QCL ($\mu_{b}=0.6 \mu_{th}$).}
\label{act}
\end{figure}
\begin{figure}[h!t]
\begin{center}
\subfigure[]{\includegraphics[width=0.47\textwidth]{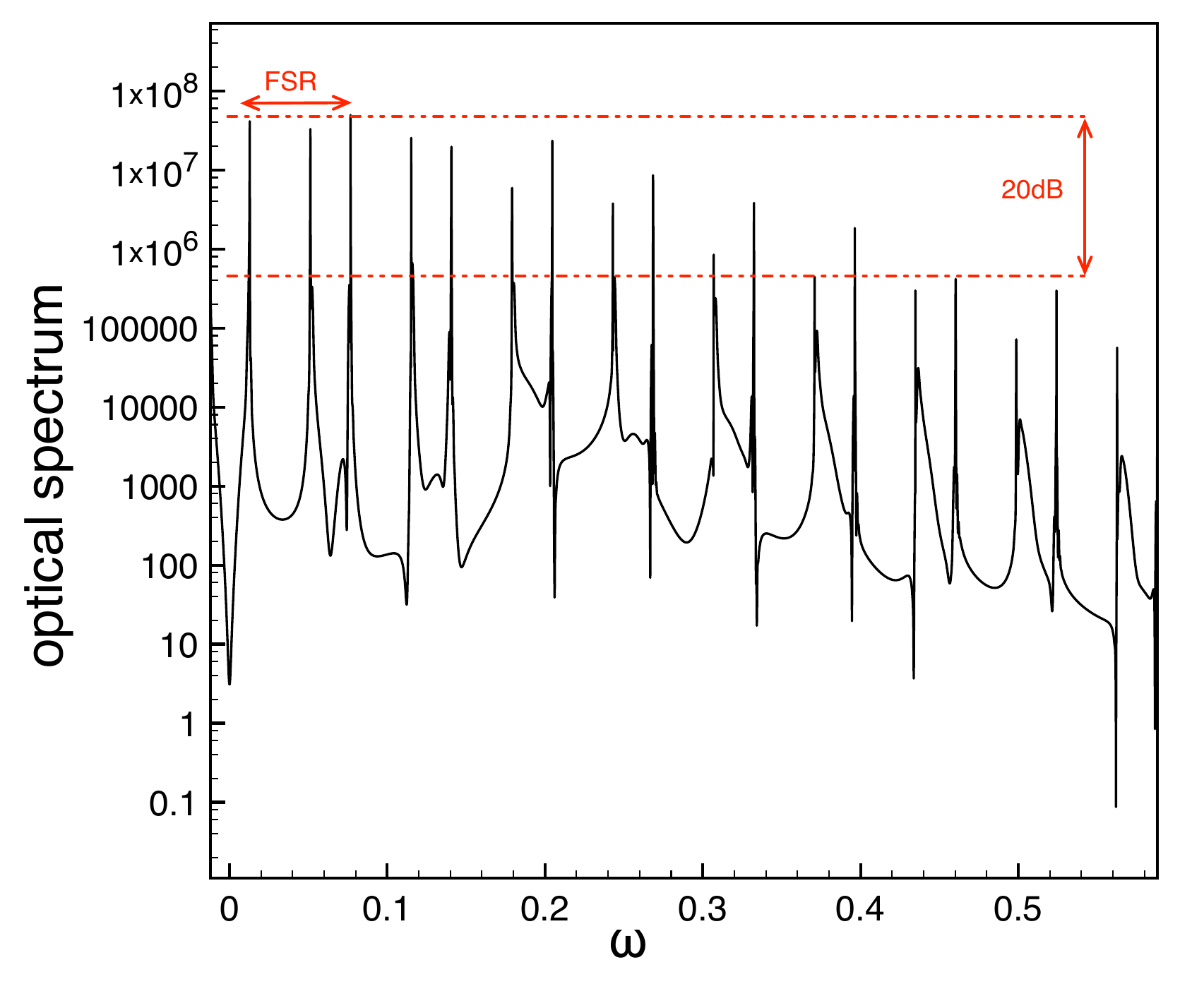}}
\subfigure[]{\includegraphics[width=0.47\textwidth]{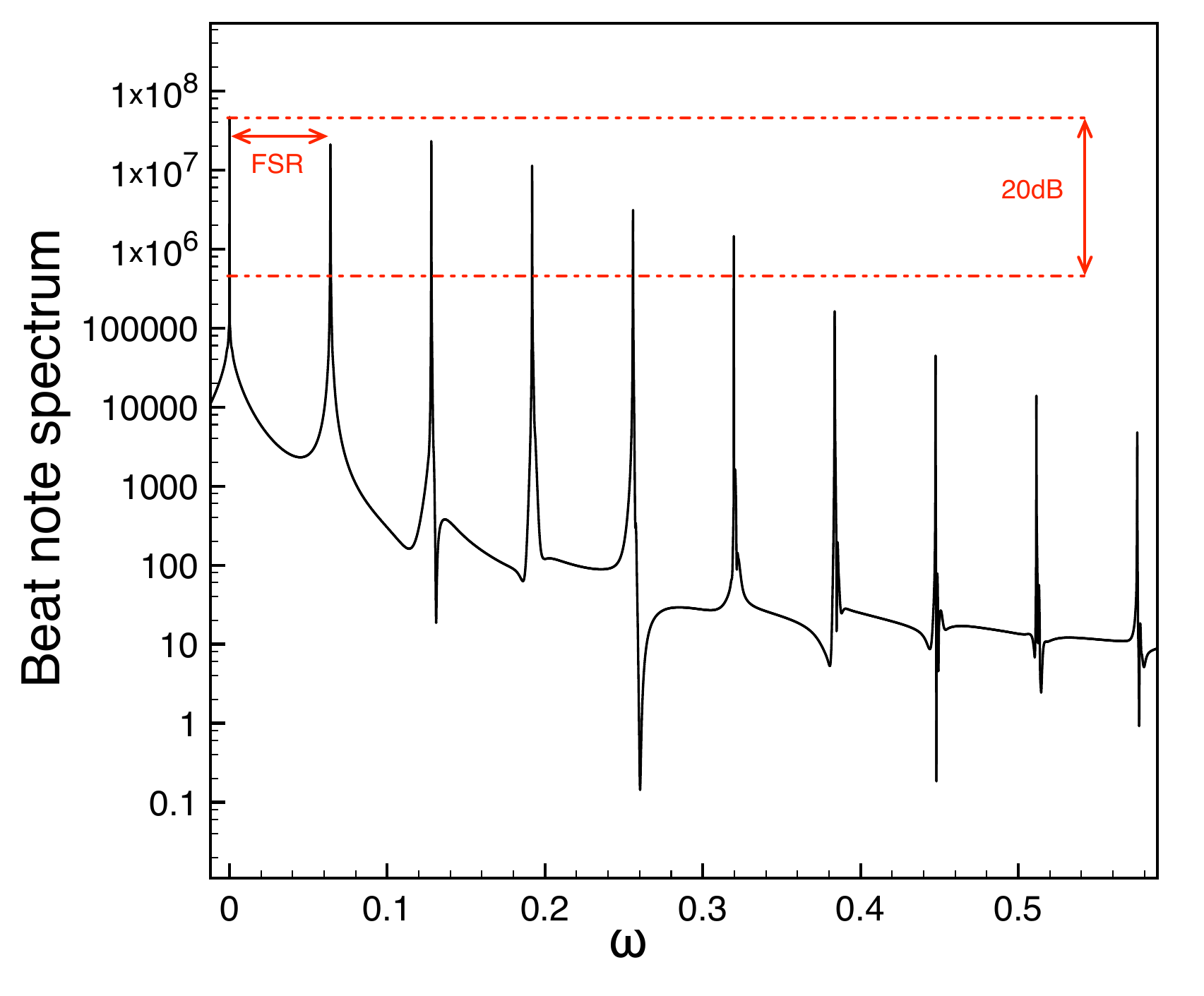}}
\end{center}
\vspace{-0.2cm}
\caption{Parameters as in Fig. \ref{act}. Optical spectrum (a) and beat note spectrum (b) corresponding to the mode-locked pulses in Fig. \ref{act}.}
\label{actspe}
\end{figure}

%
%
In Figure \ref{act} we plot the field intensity at $z=1/2$(Fig. \ref{act}.a) and the intracavity field intensity (Fig. \ref{act}.b) versus time at steady state for $\mu_{b}=0.6 \mu_{th}$, $\mu_{0}=1.5 \mu_{th}$ and $\mu_{A}=4 \mu_{th}$. Fig. \ref{act} shows the formation of pulses with FWHM of $\simeq 1.5$ $ps$, a repetition rate $\Omega_{M}$ and a contrast $S=(Max(I)-Min(I))/Max(I)=1$. The corresponding OS and BNS are shown in Fig. \ref{actspe} where the first is characterized by $13$ modes in the first two decades. Robustness was assessed by reaching the same regime with varied parameters: e.g. for a DC bias slightly above the lasing threshold ($\mu_{b}=1.1 \mu_{th}$) and we also verified that a variation of few percents in the pulse width can be obtained by varying the modulation amplitude $\mu_{A}$.\\
We checked that active mode-locking is also achieved in case of {\it faster} carriers (down to sub-picosecond time scale), and {\it broader} gain linewidth (up to $10 THz$ range). As expected, slower carriers are associated to longer pulses and above the picosecond time scale, the effect disappears because of the medium inertia.
An experimental validation of our numerical predictions would pave the way towards a number of fascinating applications based of actively mode-locked THz-QCL.

\section{Conclusion}\label{concl}
In conclusion we have established a unidirectional broad-area QCL model describing multimode regimes which focuses on features typical of active s.c. media, such as an asymmetric dispersion/gain line and a LEF. We have analytically derived an expression for the TW solutions and their lasing thresholds. While retaining only a minimum of mechanisms typical of the complex QCL device, our model proves capable of predicting low threshold multimode regimes, coherent multimode dynamics and it is capable to reproduce a number of experimentally confirmed features typical of broad-band THz-QCLs, such as the alternance between broad and narrow BN, but also of coherent regimes where we found that a narrow BN corresponds to stable self-confined structures that travel in the cavity at the group velocity of light and whose number depends on initial conditions. From a fundamental point of view, in spite of their low-contrast due to the small number of active modes, these coherent structures represent an interesting and unique phenomenon of spontaneous mode-locking in a QCL.
Also, we could validate how the effect of a radio frequency modulation of the pump facilitates a coherent dynamics and leads to active mode-locking where the laser emits picoseconds pulses. These evidences might have a serious impact on applications such as high-precision spectroscopy and time resolved measurements in the Mid-IR and THz windows of the electromagnetic spectrum.

\section*{Acknowledgments}
L. L. C and M. B thanks prof. F. Prati for many useful discussions.  L. L. C also acknowledges support from COST Action BM1205.

\appendix
\section{Appendix}
In this Appendix we provide a detailed derivation of the Effective Semiconductor Bloch Equations in the low transmission limit.\\
The boundary condition for the field envelope at $z=0=L$ are:
\begin{equation}
E(0,t)=RE(l,t-(L-l)/c)e^{ik_{0}l+i\omega_{0}(L-l)/c} \Leftrightarrow E(0,t)=RE(l,t-\Delta t)e^{i\omega_{0}\Lambda/c}
\end{equation}
with $R=1-T$ (in the hypothesis of no mirror absorption), $\Delta t=(L-l)/c$ and $\Lambda=L-l+nl$. Finally, observing that: $$e^{i\frac{\omega_{0}\Lambda}{c}}= e^{-\frac{i\Lambda}{c}\left(\frac{2 \pi m\,c}{\Lambda}-\omega_{0}\right)}=e^{-\frac{i\Lambda}{c}(\omega_{c}-\omega_{0})}=e^{-i\delta_{0}}, \quad m=0,\pm 1, \pm 2, ..$$ where $\delta_{0}$ is $2 \pi$ times the detuning between the reference angular frequency $\omega_{0}$ and the closest ring cavity resonance denoted as $\omega_{c}$ normalized on the cavity free spectral range (FSR) and that in our case ($\delta_{0}=0$), we get:
\begin{equation}
E(0,t)=RE(l,t-\Delta t) \label{BC0}
\end{equation}
If we now introduce the following transformation of independent variables \cite{NOS}:
\begin{equation}
\eta=z \quad \quad t'=t+\frac{z}{l}\Delta t
\end{equation}
the boundary condition (\ref{BC0}) assumes the isochronous form:
\begin{equation}
E(0,t')=RE(l,t') \label{BC1}
\end{equation}
Moreover, since:
$$ \frac{\partial }{\partial z}=\frac{\partial }{\partial \eta}+\frac{\Delta t}{l}\frac{\partial }{\partial t'}, \quad \quad  \frac{\partial }{\partial t}= \frac{\partial }{\partial t'}$$
Equations (\ref{E21})-(\ref{P21}) become:
 \begin{eqnarray}
\frac{\partial E}{\partial \eta}+\frac{\Lambda}{l\,c}\frac{\partial E}{\partial t'} &=& \tilde{g} P \label{E2}\\
\frac{\partial P}{\partial t'}&=& \frac{1}{\tau_{d}} \Gamma\left(1-i \alpha \right)\left[(1-i\alpha)E \, D-P \right]   \\
\frac{\partial D}{\partial t'}&=&\frac{1}{\tau_{e}}\left[ \mu-D-\frac{1}{2}(E^{*}P+E P^{*})\right] \label{P2}
\end{eqnarray}
Finally, by setting \cite{NOS}:
\begin{equation*}
E'(\eta,t')=E(\eta,t')e^{[(\ln R)\eta/l]}\label{au0}, \quad P'(\eta,t')=P(\eta,t')e^{[(\ln R)\eta/l]} 
\end{equation*}
we derive:
 \begin{eqnarray}
\frac{\partial E'}{\partial \eta}+\frac{\Lambda}{l\,c}\frac{\partial E'}{\partial t'} &=& \frac{1}{l}(\ln R) E'+\tilde{g} P' \label{E2b}\\
\frac{\partial P'}{\partial t'}&=& \frac{1}{\tau_{d}}\Gamma\left(1-i \alpha \right)\left[(1-i\alpha)DE'-P' \right]   \\
\frac{\partial D}{\partial t'}&=&\frac{1}{\tau_{e}}\left \{ \mu-D-\frac{1}{2} e^{-(2\ln R) \eta/l} \left[E'^{*}P'+ E'P'^{*} \right] \right \} \label{P2b}
\end{eqnarray}
with the periodic boundary condition:
\begin{equation}
E'(0,t')=E'(l,t')
\end{equation}
At this point, in order to simplify the theoretical analysis and in agreement with the experimental results, we assume valid the low transmission approximation defined as \cite{NOS}:
$$\tilde{g}\,l<<1, \quad \quad T<<1$$
with:
$$\mathcal{A}=\frac{\tilde{g}l}{T}=\mathcal{O}(1)$$
In this limit the auxiliary variables $E'$ and $P'$ defined by Eq. (\ref{au0}) coincide with $E$ and $P$, respectively, and the dynamical equations
(\ref{E2b})-(\ref{P2b}) reduce to:
 \begin{eqnarray}
\frac{l\,c}{\Lambda}\frac{\partial E}{\partial \eta}+\frac{\partial E}{\partial t'} &=& \frac{1}{\tau_{p}}\left[-E+\mathcal{A} P\right] \label{E31}\\
\frac{\partial P}{\partial t'}&=& \frac{1}{\tau_{d}} \Gamma\left(1-i \alpha \right)\left[(1-i\alpha)E \, D-P \right]   \label{P3a}\\
\frac{\partial D}{\partial t'}&=&\frac{1}{\tau_{e}}\left [ \mu-D-\frac{1}{2}\left(E^{*}P+ E P^{*} \right)\right]\label{P31}
\end{eqnarray}
where we introduced the photon decay time $\tau_{p}=\frac{\Lambda}{c\,T}$.\\
The boundary condition becomes:
\begin{equation}
E(0,t')=E(l,t') \label{BCnew}
\end{equation}
If we scale the time $t'$ on the fastest time scale given by the dipole de-phasing time $\tau_{d}$, and we set $\tilde{E}=E$, $\tilde{P}=\mathcal{A} P$ and $\tilde{D}=\mathcal{A} D$ we get:
 \begin{eqnarray}
\frac{l\,c \tau_{d}}{\Lambda}\frac{\partial \tilde{E}}{\partial \eta}+\frac{\partial \tilde{E}}{\partial t'} &=&\sigma \left(-\tilde{E}+\tilde{P}\right) \label{E4}\\
\frac{\partial \tilde{P}}{\partial t'}&=& \Gamma\left(1-i \alpha \right)\left[(1-i\alpha)\tilde{E} \tilde{D}-\tilde{P} \right]   \label{P4a}\\
\frac{\partial \tilde{D}}{\partial t'}&=&b\left [ \tilde{\mu}-\tilde{D}-\frac{1}{2}\left(\tilde{E}^{*}\tilde{P}+ \tilde{E} \tilde{P}^{*} \right)\right]\label{P4}
\end{eqnarray}
where $\sigma=\frac{\tau_{d}}{\tau_{p}}$ and $b=\frac{\tau_{d}}{\tau_{e}}$ and $\tilde{\mu}=\mathcal{A} \mu$\\

We observe at this point that the model is capable of describing a laser with an extended cavity since $\Lambda$, $l$ and $L$ can be chosen freely. This allows to consider in the future emitters where the photon lifetime can be controlled and the ratio of the medium and field rates may be varied to tune the competing dynamics of field, coherence and medium thus tailoring dynamical regimes of interest as in the case of bipolar VCSEL emitters \cite{Gustave}.

\end{document}